\documentclass[prx,twocolumn]{revtex4}

\usepackage{graphicx}
\usepackage{dcolumn}
\usepackage{bm}
\usepackage{multirow}
\usepackage{amsmath,amssymb}

\begin{document}


\title{Emergence and evolution of electronic modes with temperature in spin-gapped Mott and Kondo insulators}

\author{Masanori Kohno}
\email{KOHNO.Masanori@nims.go.jp}
\affiliation{Research Center for Materials Nanoarchitectonics, National Institute for Materials Science, Tsukuba 305-0003, Japan}

\date{\today}

\begin{abstract}
Electronic modes emerge within the band gap at nonzero temperature in strongly correlated insulators such as Mott and Kondo insulators, 
exhibiting momentum-shifted magnetic dispersion relations from the band edges. 
As the temperature increases, the emergent modes gain considerable spectral weights and form robust bands that differ from the zero-temperature bands. 
Here, the origin of the emergent modes, their relation to doping-induced modes, and how their spectral weights increase with temperature 
are clarified in the ladder and bilayer Hubbard models and one- and two-dimensional Kondo lattice models 
using effective theory for weak inter-unit-cell hopping and numerical calculations. 
The results indicate that the temperature-driven change in the band structure reflecting spin excitation, 
including the change in the number of bands, can be observed in various strongly correlated insulators even with a spin gap, 
provided that the temperature increases up to about spin-excitation energies. 
The controlled analyses developed in this study significantly contribute to the fundamental understanding of the band structure of strongly correlated insulators 
at nonzero temperature. 
\end{abstract}

\maketitle

\section{Introduction} 
In a conventional band insulator, wherein interactions between the electrons can be neglected, the band structure does not change with temperature. 
However, in strongly correlated insulators such as Mott and Kondo insulators, the band structure can change with temperature, 
even without a phase transition or lattice distortion. 
The origin of changes in the band structure has recently been revealed from the viewpoint of spin-charge separation of strongly correlated insulators 
by considering the selection rules and using numerical calculations \cite{KohnoMottT}. 
In short, spin excited states, whose excitation energies are lower than the charge gap, emerge within the band gap in the electronic spectrum 
at nonzero temperature, exhibiting momentum-shifted magnetic dispersion relations from the band edges \cite{KohnoMottT}. 
This feature is in contrast to a conventional band insulator, wherein electronic states reflecting spin excitation of the band insulator 
do not emerge within the band gap because spin-charge separation does not occur (the lowest spin-excitation energy is the same as the charge gap) 
(Sec. \ref{sec:spinchargeSep}). 
\par
The origin of this feature is shared with the doping-induced electronic mode whose origin has been identified with the spin excited states of the Mott and Kondo insulators; 
it has been shown that an electronic mode emerges from the top (bottom) of the lower (upper) band via hole (electron) doping, 
exhibiting the momentum-shifted magnetic dispersion relation 
\cite{Kohno1DHub,Kohno2DHub,KohnoDIS,KohnoRPP,KohnoHubLadder,KohnoKLM,KohnoGW,KohnoAF,KohnoSpin,Kohno1DtJ,Kohno2DtJ}. 
The spectral weight of the doping-induced mode has been numerically and theoretically shown to increase in proportion to the doping concentration in the small-doping regime 
\cite{Kohno1DHub,Kohno2DHub,KohnoDIS,KohnoRPP,KohnoHubLadder,KohnoKLM,KohnoGW,KohnoAF,KohnoSpin,Kohno1DtJ,Kohno2DtJ,Eskes,DagottoDOS}, 
forming a robust band that differs from the undoped bands. 
\par
The spectral weights of temperature-induced modes have been shown to increase with temperature using numerical calculations 
\cite{KohnoMottT,NoceraDMRT_T,Matsueda_QMC,PreussQP,GroberQMC,VCALanczos,TPQSLanczos,KuzminCPT,PAMMPS}. 
In this paper, to clarify how the spectral weights increase with temperature theoretically, 
we consider decoupled unit cells, whose spectral function can be calculated exactly at any temperature (Sec. \ref{sec:ucT}), 
extend the analyses to weakly coupled unit cells using the perturbation theory (Sec. \ref{sec:AkwT}), and 
verify the validity using numerical calculations for the ladder and bilayer Hubbard models and one-dimensional (1D) and two-dimensional (2D) Kondo lattice models. 
\par
The analyses and numerical results indicate that the spectral weights of temperature-induced electronic modes can become comparable to 
those of zero-temperature bands, provided that the temperature increases up to about spin-excitation energies, 
even in spin-gapped Mott and Kondo insulators 
for which one might suspect that the required temperature is about the excitation energy of the thermal state from the ground state, 
i.e., the order of the spin gap times the system size (Sec. \ref{sec:remarks}). 
\section{Models and methods} 
The Hubbard model and the Kondo lattice model (KLM) are defined by the following Hamiltonians: 
\begin{align}
{\cal H}_{\rm Hub}=&-\sum_{\langle i,j\rangle,\sigma}t_{i,j}\left(c^{\dagger}_{i,\sigma}c_{j,\sigma}+{\rm H.c.}\right)\nonumber\\
&+U\sum_i\left(n_{i,\uparrow}-\frac{1}{2}\right)\left(n_{i,\downarrow}-\frac{1}{2}\right)-\mu\sum_{i,\sigma}n_{i,\sigma},\\
{\cal H}_{\rm KLM}=&-\sum_{\langle i,j\rangle,\sigma}t_{i,j}\left(c^{\dagger}_{i,\sigma}c_{j,\sigma}+{\rm H.c.}\right)
+J_{\rm K}\sum_i{\bm S}^c_i\cdot{\bm S}^f_i\nonumber\\
&-\mu\sum_{i}\left(\sum_{\sigma}n_{i,\sigma}+1\right),
\end{align}
where $\langle i,j\rangle$ means that sites $i$ and $j$ are the nearest neighbors. 
Here, $c^{\dagger}_{i,\sigma}$ and $n_{i,\sigma}$ denote 
the creation and number operators of a conduction-orbital electron with spin $\sigma(=\uparrow,\downarrow)$ at site $i$, respectively, 
and ${\bm S}^c_i$ (${\bm S}^f_i$) denotes the spin operator of a conduction-orbital (localized-orbital) electron at site $i$. 
In the KLM, each localized orbital has one electron. 
Hereafter, $t_{i,j}=t_{\perp}$ between the chains and between the planes for the ladder and bilayer Hubbard models, respectively, and $t_{i,j}=t$ in the chains and planes. 
Unless otherwise mentioned, $t_{\perp}>0$, $U>0$, and $J_{\rm K}>0$. 
\par
The numbers of electrons, sites, and unit cells are denoted by $N_{\rm e}$, $N_{\rm s}$, and $N_{\rm u}$, respectively. 
The unit cell is a pair of sites connected by $t_{\perp}$ in the ladder and bilayer Hubbard models ($N_{\rm u}=\frac{N_{\rm s}}{2}$) and 
a site with a conduction orbital and a localized orbital in the KLM ($N_{\rm u}=N_{\rm s}$). 
The hole-doping concentration $\delta$ is defined as $\delta=1-\frac{N_{\rm e}}{2N_{\rm u}}$. At half filling, $\delta=0$, where $\mu=0$ unless otherwise mentioned. 
Hereafter, the units $\hbar=1$ and $k_{\rm B}=1$ are used. The length and $z$ component of spin are denoted as $S$ and $S^z$, respectively. 
The electronic spin opposite to $\sigma$ is denoted as ${\bar \sigma}$: ${\bar \sigma}=\downarrow$ and $\uparrow$ for $\sigma=\uparrow$ and $\downarrow$, respectively. 
The $z$ component of spin of an electron is denoted as $s^z$: $s^z=\frac{1}{2}$ and $-\frac{1}{2}$ for $\sigma=\uparrow$ and $\downarrow$, respectively. 
\par
The spectral function at temperature $T$ is defined as 
\begin{align}
A({\bm k},\omega)=\frac{1}{2\Xi}\sum_{n,m,\sigma}e^{-\beta E_n}&[|\langle m|c^{\dagger}_{{\bm k},\sigma}|n\rangle|^2\delta(\omega-E_m+E_n)\nonumber\\
+&|\langle m|c_{{\bm k},\sigma}|n\rangle|^2\delta(\omega+E_m-E_n)], 
\label{eq:Akw}
\end{align}
where $\Xi=\sum_n e^{-\beta E_n}$ and $\beta=\frac{1}{T}$. Here, $|n\rangle$ denotes the $n$th eigenstate with energy $E_n$, and 
$c^{\dagger}_{{\bm k},\sigma}$ denotes the Fourier transform of $c^{\dagger}_{i,\sigma}$. 
The spectral function can be rewritten as 
\begin{align}
A({\bm k},\omega)=\frac{1}{2\Xi}\sum_{n,m,\sigma}&(e^{-\beta E_m}+e^{-\beta E_n})|\langle m|c^{\dagger}_{{\bm k},\sigma}|n\rangle|^2\nonumber\\
&\times\delta(\omega-E_m+E_n). 
\label{eq:Akw2}
\end{align}
The spectral function can also be obtained as 
\begin{equation}
A({\bm k},\omega)=-\frac{1}{2\pi}\sum_{\sigma}{\rm Im}G_{{\bm k},\sigma}(\omega) 
\end{equation}
using the Green's function, which is expressed as 
\begin{equation}
G_{{\bm k},\sigma}(\omega)=-i\lim_{\epsilon\rightarrow+0}\int_0^{\infty}dt e^{i\omega t-\epsilon t}
\frac{\sum_l{}_l\langle\beta|\{c_{{\bm k},\sigma}(t),c^{\dagger}_{{\bm k},\sigma}\}|\beta\rangle_l}{\sum_l{}_l\langle\beta|\beta\rangle_l},
\end{equation}
where 
\begin{equation}
c_{{\bm k},\sigma}(t)=e^{i{\cal H}t}c_{{\bm k},\sigma}e^{-i{\cal H}t},\quad |\beta\rangle_l=e^{-\frac{\beta{\cal H}}{2}}|{\rm I}\rangle_l. 
\label{eq:thermalState}
\end{equation}
Here, ${\cal H}$ represents ${\cal H}_{\rm Hub}$ and ${\cal H}_{\rm KLM}$. 
\par
The momentum ${\bm k}$ for the ladder (bilayer) Hubbard model has a leg (in-plane) component ${\bm k}_{\parallel}$ and a rung (out-of-plane) component $k_{\perp}$: 
${\bm k}_{\parallel}=k_x$ and $k_{\perp}=k_y$ in the ladder Hubbard model; 
${\bm k}_{\parallel}=(k_x,k_y)$ and $k_{\perp}=k_z$ in the bilayer Hubbard model. In the KLM, ${\bm k}_{\parallel}={\bm k}$. 
\par
The dynamical spin structure factors of the Hubbard model and KLM at zero temperature are defined as follows: 
\begin{equation}
\begin{array}{l}
S_{\rm Hub}({\bm k}_{\parallel},\omega)=\sum_{n,\alpha}|\langle n|S^{c,\alpha}_{({\bm k}_{\parallel},\pi)}|{\rm GS}\rangle|^2\delta(\omega-e_n),\\
S_{\rm KLM}({\bm k},\omega)=\frac{1}{2}\sum_{n,\alpha}|\langle n|(S^{c,\alpha}_{\bm k}-S^{f,\alpha}_{\bm k})|{\rm GS}\rangle|^2\delta(\omega-e_n), 
\end{array}
\label{eq:Skw}
\end{equation}
where $S^{\lambda,\alpha}_{\bm k}$ denotes the Fourier transform of $S^{\lambda,\alpha}_i$ for $\lambda=c$ and $f$; 
$S^{\lambda,\alpha}_i$ denotes the $\alpha$-component of ${\bm S}^{\lambda}_i$ for $\alpha=x$, $y$, and $z$. 
Here, $e_n$ denotes the excitation energy of $|n\rangle$ from the ground state $|{\rm GS}\rangle$. 
\par
The dynamical charge structure factor of the Hubbard model at zero temperature is defined as 
\begin{equation}
N_{\rm Hub}({\bm k}_{\parallel},\omega)=\sum_{m}|\langle m|{\tilde n}_{({\bm k}_{\parallel},\pi)}|{\rm GS}\rangle|^2\delta(\omega-e_m),
\label{eq:Nkw}
\end{equation}
where ${\tilde n}_{\bm k}=\frac{1}{\sqrt{N_{\rm s}}}\sum_{i}e^{i{\bm k}\cdot{\bm r}_i}(n_i-{\bar n})$ 
with $n_i=\sum_{\sigma}n_{i,\sigma}$ and ${\bar n}=\frac{N_{\rm e}}{N_{\rm s}}$. 
Here, $e_m$ denotes the excitation energy of $|m\rangle$ from the ground state $|{\rm GS}\rangle$. 
\par
In this paper, the spectral functions at nonzero temperatures calculated using the cluster perturbation theory (CPT) \cite{CPTPRL,CPTPRB} are shown. 
In the CPT, the real-space Green's functions of the 1D and 2D KLMs were calculated using the full exact-diagonalization method, and 
those of the ladder and bilayer Hubbard models were calculated using the low-temperature Lanczos method (LTLM) \cite{LTLM}, 
which is equivalent to the thermal-pure-quantum-state method with the Lanczos solver \cite{TPQS,TPQSLanczos}. 
In the LTLM, typically 10 random vectors were generated for $|{\rm I}\rangle_l$ in Eq. (\ref{eq:thermalState}), 
and orthonormal states obtained from the random vectors via QP decomposition were used as the initial vectors of the block Lanczos method. 
Typically 600 eigenstates obtained by the block Lanczos method were used to construct the thermal states $|\beta\rangle_l$ [Eq. (\ref{eq:thermalState})] 
in each subspace specified by the numbers of up and down spins. 
The Green's functions were calculated in 6-unit-cell clusters ($3\times 2$-unit-cell clusters for the bilayer Hubbard model and 2D KLM). 
The numerical results for the 1D and 2D Hubbard models obtained using the above method \cite{KohnoMottT} are consistent 
with those obtained using the time-dependent density-matrix renormalization group method \cite{NoceraDMRT_T} 
and quantum Monte Carlo method \cite{PreussQP,GroberQMC}, respectively. 
\par
The spectral function and the dynamical spin and charge structure factors at zero temperature were calculated 
using the non-Abelian dynamical density-matrix renormalization group (DDMRG) method \cite{KohnoDIS,Kohno1DtJ,Kohno2DtJ,KohnoHubLadder,nonAbelianHub,nonAbeliantJ,nonAbelianThesis,DDMRG} 
in the ladder Hubbard model for $U>0$ and the 1D KLM. 
The non-Abelian DDMRG calculations were performed under open boundary conditions on clusters with $N_{\rm u}=60$, 
wherein 120 eigenstates of the density matrix were retained. 
The spectral function of the bilayer Hubbard model (2D KLM) at zero temperature was calculated using the CPT on $3\times 2$-unit-cell ($3\times 3$-site) clusters 
with the continued fraction expansion in the Lanczos method. 
The lowest spin-excitation energy at each ${\bm k}$ point in the bilayer Hubbard model and 2D KLM was calculated under periodic boundary conditions 
using the Lanczos method. 
\par
The symmetrized spectral function ${\bar A}({\bm k},\omega)$ is defined as $[A(k_x,k_y,k_{\perp},\omega)+A(k_y,k_x,k_{\perp},\omega)]/2$ and 
$[A(k_x,k_y,\omega)+A(k_y,k_x,\omega)]/2$ for the bilayer Hubbard model and 2D KLM, respectively. 
The spectral weight of a band is defined as $\frac{1}{N_{\rm u}}\sum_{\bm k}\int d\omega A_{\rm band}({\bm k},\omega)$, 
where $A_{\rm band}({\bm k},\omega)$ denotes the spectral function of the band. 
\section{Results and discussion} 
\subsection{Overall spectral features} 
\begin{figure*}
\includegraphics[width=\linewidth]{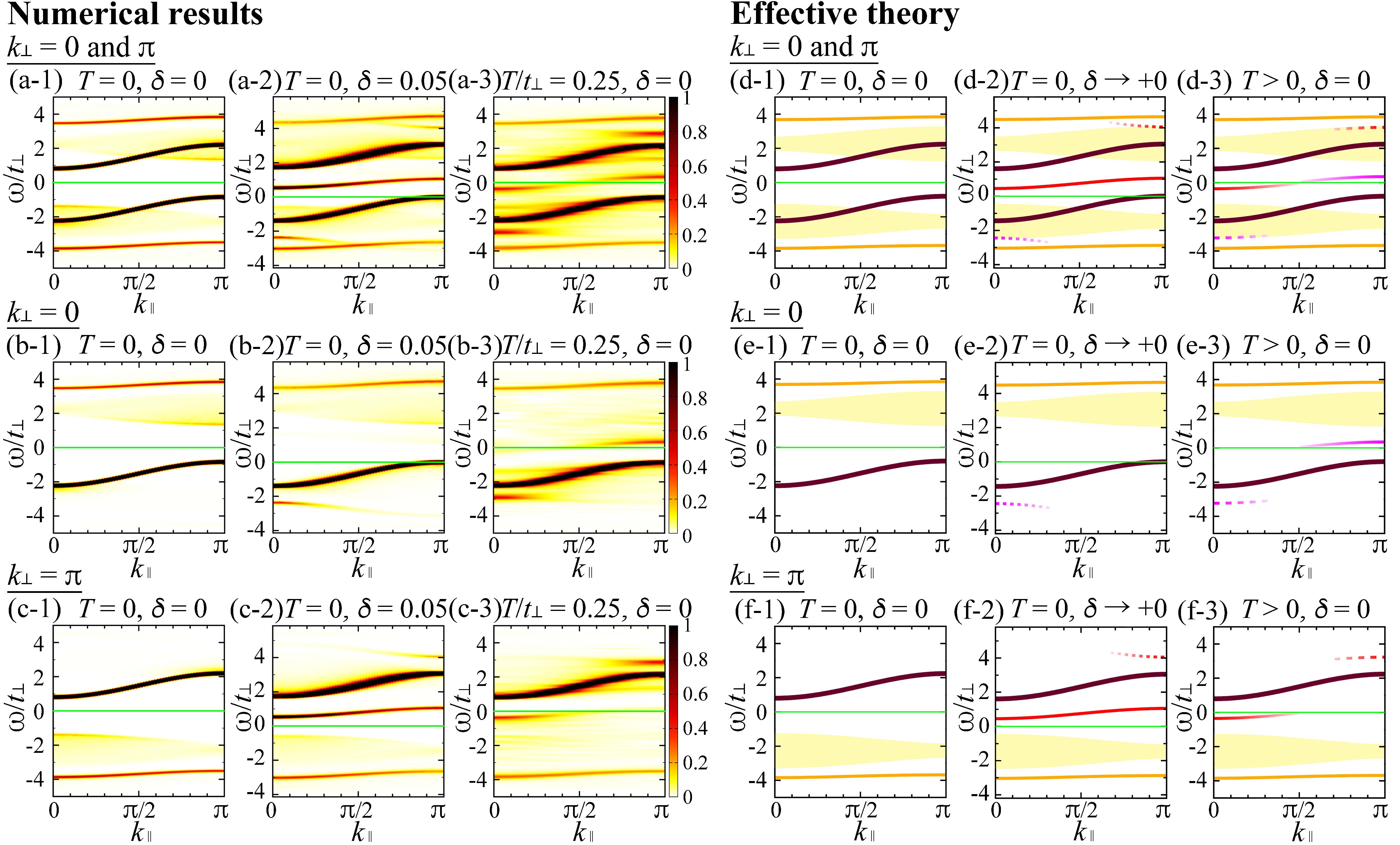}
\caption{Electronic excitation of the ladder Hubbard model for $t/t_{\perp}=0.4$ and $U/t_{\perp}=3$. 
(a-1)--(a-3) $A(k_{\parallel},0,\omega)t_{\perp}+A(k_{\parallel},\pi,\omega)t_{\perp}$, 
(b-1)--(b-3) $A(k_{\parallel},0,\omega)t_{\perp}$, and (c-1)--(c-3) $A(k_{\parallel},\pi,\omega)t_{\perp}$ 
obtained using the non-Abelian DDMRG method for $T=0$ at $\delta=0$ with $\mu=0$ [(a-1)--(c-1)]; 
$T=0$ at $\delta=0.05$ with $\mu$ adjusted for $\delta=0.05$ [(a-1)--(c-2)] 
and obtained using the CPT for $T/t_{\perp}=0.25$ at $\delta=0$ with $\mu=0$ [(a-3)--(c-3)]. 
Gaussian broadening with a standard deviation of $0.05t_{\perp}$ is used. 
(d-1)--(d-3), (e-1)--(e-3), (f-1)--(f-3) Dispersion relations obtained using the effective theory for weak inter-unit-cell hopping 
for $T=0$ at $\delta=0$ with $\mu=0$ [(d-1)--(f-1)]; $T=0$ at $\delta\rightarrow +0$ with $\mu$ of $\varepsilon^{\rm B}_{-\pi}=0$ [(d-2)--(f-2)]; 
$T>0$ at $\delta=0$ with $\mu=0$ [(d-3)--(f-3)]. 
The curves represent $\omega=-\varepsilon^{\rm B}_{-k_{\parallel}}$ (solid brown curves for $\omega\le 0$) [(d-1)--(d-3), (e-1)--(e-3)], 
$\omega=\varepsilon^{\rm F}_{k_{\parallel}}$ (solid brown curves for $\omega>0$) [(d-1)--(d-3), (f-1)--(f-3)], 
$\omega=\varepsilon^{\rm G}_{k_{\parallel}}$ (solid orange curves for $\omega>0$) [(d-1)--(d-3), (e-1)--(e-3)], 
$\omega=-\varepsilon^{\rm A}_{-k_{\parallel}}$ (solid orange curves for $\omega<0$) [(d-1)--(d-3), (f-1)--(f-3)], 
$\omega=e^{\rm spin}_{k_{\parallel}-\pi}$ (solid red curves) [(d-2), (f-2)], 
$\omega=e^{\rm spin}_{k_{\parallel}-\pi}+\mu_-$ (solid red curves for $k_{\parallel}\approx 0$) [(d-3), (f-3)], 
$\omega=-e^{\rm spin}_{-k_{\parallel}}+\mu_+$ (solid magenta curves for $k_{\parallel}\approx\pi$) [(d-3), (e-3)]. 
$\omega=e^{{\rm D}^-}_{k_{\parallel}-\pi}$ (dotted red curves for $k_{\parallel}\approx\pi$ and $\omega>0$) [(d-2), (f-2)], 
$\omega=-e^{\rm V}_{-k_{\parallel}-\pi}$ (dotted magenta curves for $k_{\parallel}\approx 0$ and $\omega<0$) [(d-2), (e-2)], 
$\omega=-e^{\rm spin}_{-k_{\parallel}}+\varepsilon^{\rm G}_{0}$ (dashed red curves for $k_{\parallel}\approx\pi$ and $\omega>0$) [(d-3), (f-3)], and 
$\omega=e^{\rm spin}_{k_{\parallel}-\pi}-\varepsilon^{\rm A}_{-\pi}$ (dashed magenta curves for $k_{\parallel}\approx 0$ and $\omega<0$) [(d-3), (e-3)]. 
The light-yellow regions indicate $\omega=\varepsilon^{\rm F}_{k_{\parallel}-p_{\parallel}}+e^{\rm spin}_{p_{\parallel}}$ for $\omega>0$ [(d-1)--(d-3), (e-1)--(e-3)] and 
$\omega=-\varepsilon^{\rm B}_{-k_{\parallel}+p_{\parallel}}-e^{\rm spin}_{-p_{\parallel}}$ for $\omega<0$ [(d-1)--(d-3), (f-1)--(f-3)]. 
The solid green lines indicate $\omega=0$.}
\label{fig:HubLadder}
\end{figure*}
\begin{figure*}
\includegraphics[width=\linewidth]{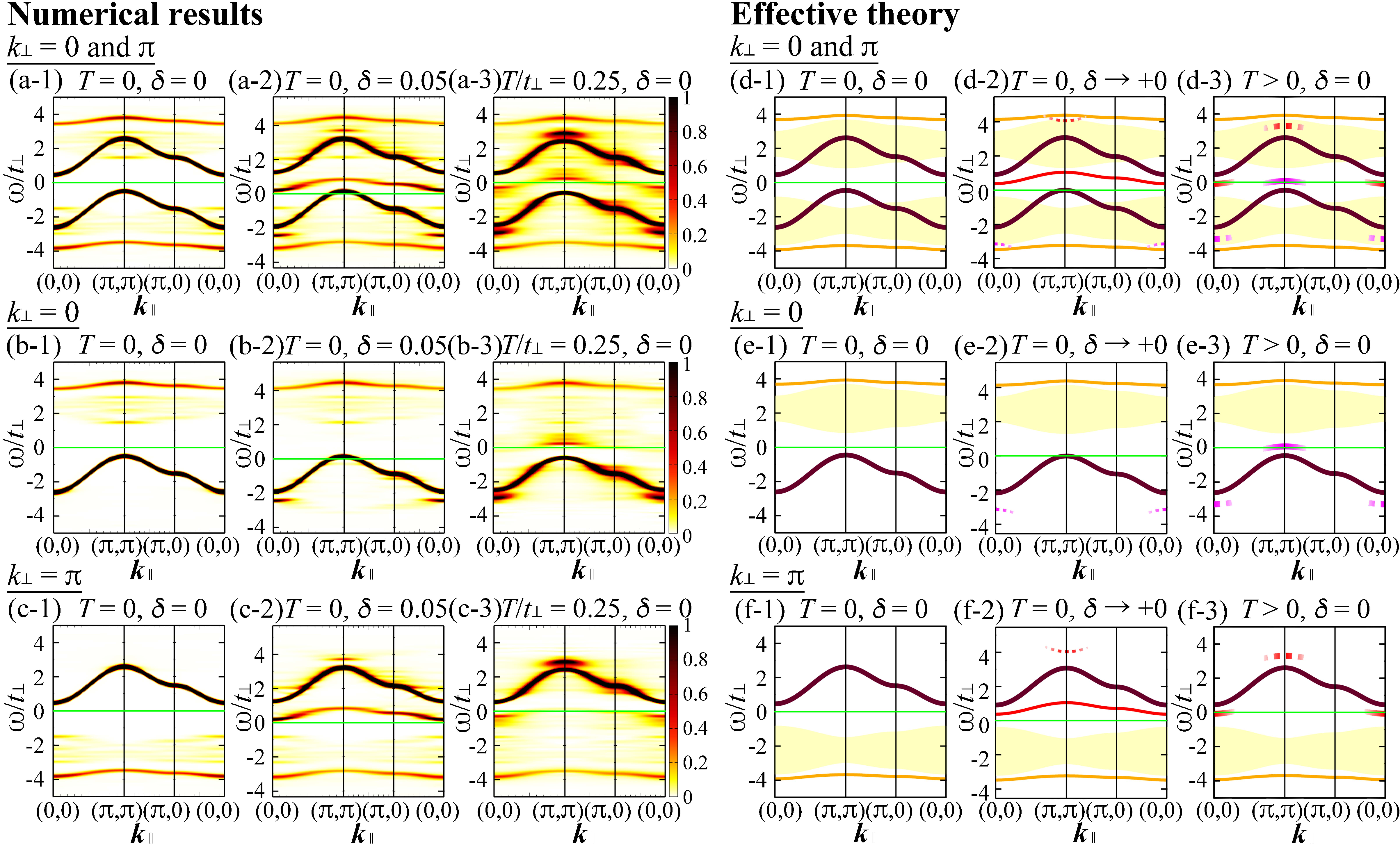}
\caption{Electronic excitation of the bilayer Hubbard model for $t/t_{\perp}=0.3$ and $U/t_{\perp}=3$. 
(a-1)--(a-3) ${\bar A}({\bm k}_{\parallel},0,\omega)t_{\perp}+{\bar A}({\bm k}_{\parallel},\pi,\omega)t_{\perp}$, 
(b-1)--(b-3) ${\bar A}({\bm k}_{\parallel},0,\omega)t_{\perp}$, (c-1)--(c-3) ${\bar A}({\bm k}_{\parallel},\pi,\omega)t_{\perp}$ 
obtained using the CPT for $T=0$ at $\delta=0$ with $\mu=0$ [(a-1)--(c-1)]; $T=0$ at $\delta=0.05$ with $\mu$ adjusted for $\delta=0.05$ [(a-2)--(c-2)]; 
$T/t_{\perp}=0.25$ at $\delta=0$ with $\mu=0$ [(a-3)--(c-3)]. 
Gaussian broadening with a standard deviation of $0.05t_{\perp}$ is used. 
(d-1)--(d-3), (e-1)--(e-3), (f-1)--(f-3) Dispersion relations obtained using the effective theory for weak inter-unit-cell hopping 
for $T=0$ at $\delta=0$ with $\mu=0$ [(d-1)--(f-1)]; $T=0$ at $\delta\rightarrow +0$ with $\mu$ of $\varepsilon^{\rm B}_{-{\bm \pi}}=0$ [(d-2)--(f-2)]; 
$T>0$ at $\delta=0$ with $\mu=0$ [(d-3)--(f-3)]. 
The curves represent $\omega=-\varepsilon^{\rm B}_{-{\bm k}_{\parallel}}$ (solid brown curves for $\omega\le 0$) [(d-1)--(d-3), (e-1)--(e-3)], 
$\omega=\varepsilon^{\rm F}_{{\bm k}_{\parallel}}$ (solid brown curves for $\omega>0$) [(d-1)--(d-3), (f-1)--(f-3)], 
$\omega=\varepsilon^{\rm G}_{{\bm k}_{\parallel}}$ (solid orange curves for $\omega>0$) [(d-1)--(d-3), (e-1)--(e-3)], 
$\omega=-\varepsilon^{\rm A}_{-{\bm k}_{\parallel}}$ (solid orange curves for $\omega<0$) [(d-1)--(d-3), (f-1)--(f-3)], 
$\omega=e^{\rm spin}_{{\bm k}_{\parallel}-{\bm \pi}}$ (solid red curves) [(d-2), (f-2)], 
$\omega=e^{\rm spin}_{{\bm k}_{\parallel}-{\bm \pi}}+\mu_-$ (solid red curves for ${\bm k}_{\parallel}\approx{\bm 0}$) [(d-3), (f-3)], 
$\omega=-e^{\rm spin}_{-{\bm k}_{\parallel}}+\mu_+$ (solid magenta curves for ${\bm k}_{\parallel}\approx{\bm \pi}$) [(d-3), (e-3)]. 
$\omega=e^{{\rm D}^-}_{{\bm k}_{\parallel}-{\bm \pi}}$ (dotted red curves for ${\bm k}_{\parallel}\approx{\bm \pi}$ and $\omega>0$) [(d-2), (f-2)], 
$\omega=-e^{\rm V}_{-{\bm k}_{\parallel}-{\bm \pi}}$ (dotted magenta curves for ${\bm k}_{\parallel}\approx{\bm 0}$ and $\omega<0$) [(d-2), (e-2)], 
$\omega=-e^{\rm spin}_{-{\bm k}_{\parallel}}+\varepsilon^{\rm G}_{\bm 0}$ (dashed red curves for ${\bm k}_{\parallel}\approx{\bm \pi}$ and $\omega>0$) [(d-3), (f-3)], and 
$\omega=e^{\rm spin}_{{\bm k}_{\parallel}-{\bm \pi}}-\varepsilon^{\rm A}_{-{\bm \pi}}$ 
(dashed magenta curves for ${\bm k}_{\parallel}\approx{\bm 0}$ and $\omega<0$) [(d-3), (e-3)]. 
The light-yellow regions indicate $\omega=\varepsilon^{\rm F}_{{\bm k}_{\parallel}-{\bm p}_{\parallel}}+e^{\rm spin}_{{\bm p}_{\parallel}}$ for $\omega>0$ [(d-1)--(d-3), (e-1)--(e-3)] 
and $\omega=-\varepsilon^{\rm B}_{-{\bm k}_{\parallel}+{\bm p}_{\parallel}}-e^{\rm spin}_{-{\bm p}_{\parallel}}$ for $\omega<0$ [(d-1)--(d-3), (f-1)--(f-3)]. 
The solid green lines indicate $\omega=0$.}
\label{fig:HubBilayer}
\end{figure*}
\begin{figure*}
\includegraphics[width=\linewidth]{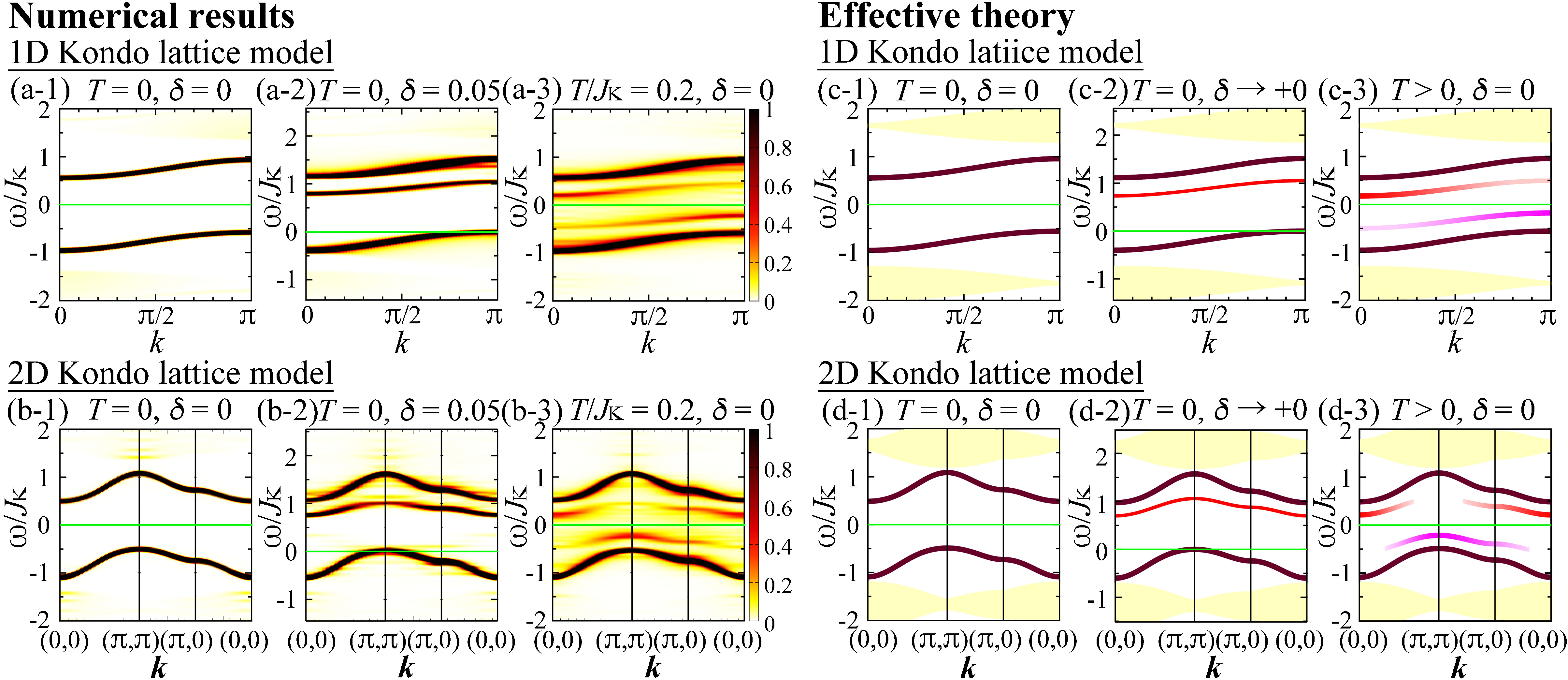}
\caption{Electronic excitation of the 1D Kondo lattice model for $t/J_{\rm K}=0.2$ [(a-1)--(a-3), (c-1)--(c-3)] and 
the 2D Kondo lattice model for $t/J_{\rm K}=0.15$ [(b-1)--(b-3), (d-1)--(d-3)]. 
(a-1)--(a-3) $A(k,\omega)J_{\rm K}$ obtained using the non-Abelian DDMRG method for $T=0$ at $\delta=0$ with $\mu=0$ [(a-1)] 
and $\delta=0.05$ with $\mu$ adjusted for $\delta=0.05$ [(a-2)] 
and obtained using the CPT for $T/J_{\rm K}=0.2$ at $\delta=0$ with $\mu=0$ [(a-3)]. 
(b-1)--(b-3) ${\bar A}({\bm k},\omega)J_{\rm K}$ obtained using the CPT for $T=0$ at $\delta=0$ with $\mu=0$ [(b-1)]; 
$T=0$ at $\delta=0.05$ with $\mu$ adjusted for $\delta=0.05$ [(b-2)]; $T/J_{\rm K}=0.2$ at $\delta=0$ with $\mu=0$ [(b-3)]. 
Gaussian broadening with a standard deviation of $0.02J_{\rm K}$ is used. 
(c-1)--(c-3), (d-1)--(d-3) Dispersion relations obtained using the effective theory for weak inter-unit-cell hopping 
for $T=0$ at $\delta=0$ with $\mu=0$ [(c-1), (d-1)]; $T=0$ at $\delta\rightarrow +0$ with $\mu$ of $\varepsilon^{\rm R}_{-{\bm \pi}}=0$ [(c-2), (d-2)]; 
$T>0$ at $\delta=0$ with $\mu=0$ [(c-3), (d-3)]. 
The curves represent $\omega=-\varepsilon^{\rm R}_{-{\bm k}}$ (solid brown curves for $\omega\le 0$) [(c-1)--(c-3), (d-1)--(d-3)], 
$\omega=\varepsilon^{\rm P}_{\bm k}$ (solid brown curves for $\omega>0$) [(c-1)--(c-3), (d-1)--(d-3)], 
$\omega=e^{\rm spin}_{{\bm k}-{\bm \pi}}$ (solid red curves) [(c-2), (d-2)], 
$\omega=e^{\rm spin}_{{\bm k}-{\bm \pi}}+\mu_-$ (solid red curves for ${\bm k}\approx{\bm 0}$) [(c-3), (d-3)], and 
$\omega=-e^{\rm spin}_{-{\bm k}}+\mu_+$ (solid magenta curves for ${\bm k}\approx{\bm \pi}$) [(c-3), (d-3)]. 
The light-yellow regions indicate $\omega=\varepsilon^{\rm P}_{{\bm k}-{\bm p}}+e^{\rm spin}_{\bm p}$ for $\omega>0$ and 
$\omega=-\varepsilon^{\rm R}_{-{\bm k}+{\bm p}}-e^{\rm spin}_{-{\bm p}}$ for $\omega<0$ [(c-1)--(c-3), (d-1)--(d-3)]. 
The solid green lines indicate $\omega=0$.}
\label{fig:KLM}
\end{figure*}
At zero temperature, the ladder and bilayer Hubbard models at half filling are Mott insulators. 
There are two dominant bands [Figs. \ref{fig:HubLadder}(a-1) and \ref{fig:HubBilayer}(a-1)]: 
one for $\omega<0$ at $k_{\perp}=0$ [Figs. \ref{fig:HubLadder}(b-1) and \ref{fig:HubBilayer}(b-1)] 
and the other for $\omega>0$ at $k_{\perp}=\pi$ [Figs. \ref{fig:HubLadder}(c-1) and \ref{fig:HubBilayer}(c-1)]. 
The band gap is defined as the gap between these two bands, which is also called the Mott gap or Hubbard gap \cite{HubbardI}. 
The 1D and 2D KLMs at half filling are Kondo insulators. 
There are two dominant bands: one for $\omega<0$ and the other for $\omega>0$ [Figs. \ref{fig:KLM}(a-1) and \ref{fig:KLM}(b-1)]. 
The band gap is defined as the gap between these two bands. 
The band gap is generally different from the spin gap in Mott and Kondo insulators \cite{Anderson,TsunetsuguRMP} (Sec. \ref{sec:spinchargeSep}). 
\par
When the Mott and Kondo insulators are doped with holes at zero temperature, the Fermi level enters the lower band, 
and an electronic mode emerges within the band gap [Figs. \ref{fig:HubLadder}(a-2), \ref{fig:HubLadder}(c-2), \ref{fig:HubBilayer}(a-2), \ref{fig:HubBilayer}(c-2), 
\ref{fig:KLM}(a-2), and \ref{fig:KLM}(b-2)], which exhibits the magnetic dispersion relation shifted by the Fermi momentum 
\cite{Kohno1DHub,Kohno2DHub,KohnoDIS,KohnoRPP,KohnoHubLadder,KohnoKLM,KohnoGW,KohnoAF,KohnoSpin,Kohno1DtJ,Kohno2DtJ}. 
\par
Similarly, at nonzero temperature, electronic modes emerge within the band gap [Figs. \ref{fig:HubLadder}(a-3)--\ref{fig:HubLadder}(c-3), 
\ref{fig:HubBilayer}(a-3)--\ref{fig:HubBilayer}(c-3), \ref{fig:KLM}(a-3), and \ref{fig:KLM}(b-3)], 
which also exhibit the magnetic dispersion relations shifted by the Fermi momenta of the small-doping limit from the band edges \cite{KohnoMottT}. 
\par
The dispersion relations of the characteristic modes can be explained reasonably well by the effective theory for weak inter-unit-cell hopping 
[Figs. \ref{fig:HubLadder}(d-1)--\ref{fig:HubLadder}(d-3), \ref{fig:HubLadder}(e-1)--\ref{fig:HubLadder}(e-3), \ref{fig:HubLadder}(f-1)--\ref{fig:HubLadder}(f-3), 
\ref{fig:HubBilayer}(d-1)--\ref{fig:HubBilayer}(d-3), \ref{fig:HubBilayer}(e-1)--\ref{fig:HubBilayer}(e-3), \ref{fig:HubBilayer}(f-1)--\ref{fig:HubBilayer}(f-3), 
\ref{fig:KLM}(c-1)--\ref{fig:KLM}(c-3), and \ref{fig:KLM}(d-1)--\ref{fig:KLM}(d-3)]. 
\par
In the following sections, the origin of the emergent modes is clarified using the effective theory (Secs. \ref{sec:effStates}--\ref{sec:TindStates}), 
and how the spectral weights of the temperature-induced states become significant with temperature is discussed 
by extending the effective theory to nonzero temperature (Secs. \ref{sec:ucT}--\ref{sec:AkwT}). 
\subsection{Effective states of weakly coupled unit cells} 
\label{sec:effStates}
At $t=0$, the system is decoupled into unit cells. 
The eigenstates of the Hubbard model and KLM in a unit cell are obtained 
as presented in Tables \ref{tbl:Hub} and \ref{tbl:KLM}, respectively 
\cite{KohnoHubLadder,KohnoKLM,OneSiteKondo,localRungHubLadder,TsunetsuguRMP}. 
The spin indices are omitted for spin-independent energies, 
e.g., $E_{\rm T}$, $\xi_{\psi^-{\rm B}}, \varepsilon^{\rm G}_{{\bm k}_{\parallel}}$, and $t_{\rm eff}^{\rm P}$. 
\begin{table} 
\caption{Eigenstates of the dimer Hubbard model}
\begin{tabular}{ccccc}
\hline\hline
$N_{\rm e}$&$S$&$k_{\perp}$&Eigenstate&Energy\\\hline
\multirow{1}{*}{4}&0&0&$|{\rm Z}\rangle=|\uparrow\downarrow,\uparrow\downarrow\rangle$&$E_{\rm Z}=\frac{U}{2}-4\mu$\\\hline 
\multirow{2}{*}{3}&\multirow{2}{*}{$\frac{1}{2}$}&$\pi$&$|{\rm F}_{\sigma}\rangle=\frac{1}{\sqrt{2}}\left(|\uparrow\downarrow,\sigma\rangle-|\sigma,\uparrow\downarrow\rangle\right)$
&$E_{\rm F}=-t_{\perp}-3\mu$\\ 
&&0&$|{\rm G}_{\sigma}\rangle=\frac{1}{\sqrt{2}}\left(|\uparrow\downarrow,\sigma\rangle+|\sigma,\uparrow\downarrow\rangle\right)$
&$E_{\rm G}=t_{\perp}-3\mu$\\\hline 
\multirow{5}{*}{2}&\multirow{2}{*}{1}&\multirow{2}{*}{$\pi$}&$|{\rm T}^{2s^z}\rangle=|\sigma,\sigma\rangle$&\multirow{2}{*}{$E_{\rm T}=-\frac{U}{2}-2\mu$}\\ 
&&&$|{\rm T}^0\rangle=\frac{1}{\sqrt{2}}\left(|\uparrow,\downarrow\rangle+|\downarrow,\uparrow\rangle\right)$&\\\cline{2-5} 
&\multirow{3}{*}{0}&0&$|\psi^-\rangle=u^+|{\rm S}\rangle+u^-|{\rm D}^+\rangle$&$E_{\psi^-}=E^--2\mu$\\ 
&&0&$|\psi^+\rangle=-u^-|{\rm S}\rangle+u^+|{\rm D}^+\rangle$&$E_{\psi^+}=E^+-2\mu$\\ 
&&$\pi$&$|{\rm D}^-\rangle$&$E_{{\rm D}^-}=\frac{U}{2}-2\mu$\\\hline 
\multirow{2}{*}{1}&\multirow{2}{*}{$\frac{1}{2}$}&$\pi$&$|{\rm A}_{\sigma}\rangle=\frac{1}{\sqrt{2}}\left(|\sigma,0\rangle-|0,\sigma\rangle\right)$&$E_{\rm A}=t_{\perp}-\mu$\\ 
&&0&$|{\rm B}_{\sigma}\rangle=\frac{1}{\sqrt{2}}\left(|\sigma,0\rangle+|0,\sigma\rangle\right)$&$E_{\rm B}=-t_{\perp}-\mu$\\\hline 
\multirow{1}{*}{0}&0&0&$|{\rm V}\rangle=|0,0\rangle$&$E_{\rm V}=\frac{U}{2}$\\\hline\hline 
\end{tabular}
\footnote[0]{\begin{tabular}{cc}
\multicolumn{2}{l}{The state with $|\alpha_i\rangle$ at site $i(=1,2)$ in a dimer is denoted as $|\alpha_1,\alpha_2\rangle$.}\\
$|{\rm S}\rangle=\frac{1}{\sqrt{2}}\left(|\uparrow,\downarrow\rangle-|\downarrow,\uparrow\rangle\right)$,
&$|{\rm D}^{\pm}\rangle=\frac{1}{\sqrt{2}}\left(|\uparrow\downarrow,0\rangle\pm|0,\uparrow\downarrow\rangle\right)$,\\
$E^{\pm}=\pm\frac{\sqrt{U^2+16t_{\perp}^2}}{2}$, 
&$u^{\pm}=\sqrt{\frac{1}{2}\left(1\pm\frac{U}{\sqrt{U^2+16t_{\perp}^2}}\right)}$.\\
\multicolumn{2}{l}{$s^z=\frac{1}{2}$ and $-\frac{1}{2}$ for $\sigma=\uparrow$ and $\downarrow$, respectively.}
\end{tabular}}
\label{tbl:Hub}
\end{table}
\begin{table} 
\caption{Eigenstates of the single-site Kondo lattice model}
\begin{tabular}{cccc}
\hline\hline
$N_{\rm e}$&$S$&Eigenstate&Energy\\\hline
\multirow{1}{*}{3}&\multirow{1}{*}{$\frac{1}{2}$}&$|{\rm P}_{\sigma}\rangle=|\uparrow\downarrow,\sigma\rangle$&$E_{\rm P}=-3\mu$\\\hline 
\multirow{3}{*}{2}&\multirow{2}{*}{1}&$|{\rm T}^{2s^z}\rangle=|\sigma,\sigma\rangle$&\multirow{2}{*}{$E_{\rm T}=\frac{J_{\rm K}}{4}-2\mu$}\\ 
&&$|{\rm T}^0\rangle=\frac{1}{\sqrt{2}}\left(|\uparrow,\downarrow\rangle+|\downarrow,\uparrow\rangle\right)$&\\\cline{2-4} 
&\multirow{1}{*}{0}&$|{\rm S}\rangle=\frac{1}{\sqrt{2}}\left(|\uparrow,\downarrow\rangle-|\downarrow,\uparrow\rangle\right)$&$E_{\rm S}=-\frac{3J_{\rm K}}{4}-2\mu$\\\hline 
\multirow{1}{*}{1}&\multirow{1}{*}{$\frac{1}{2}$}&$|{\rm R}_{\sigma}\rangle=|0,\sigma\rangle$&$E_{\rm R}=-\mu$\\\hline\hline 
\end{tabular}
\footnote[0]{\begin{tabular}{l}
The state with $|\alpha_c\rangle$ in the conduction orbital and \\
$|\alpha_f\rangle$ in the localized orbital is denoted as $|\alpha_c,\alpha_f\rangle$. \\
$s^z=\frac{1}{2}$ and $-\frac{1}{2}$ for $\sigma=\uparrow$ and $\downarrow$, respectively.
\end{tabular}}
\label{tbl:KLM}
\end{table}
\begin{table} 
\caption{Symbols of states in a unit cell}
\begin{tabular}{cccc}
\hline\hline
States&Symbols&Hubbard model&KLM\\\hline
Ground state&${\cal G}$&$\psi^-$&${\rm S}$\\ 
Low-energy state with $N_{\rm e}=1$&${\cal R}_{\sigma}$&${\rm B}_{\sigma}$&${\rm R}_{\sigma}$\\ 
Low-energy state with $N_{\rm e}=3$&${\cal A}_{\sigma}$&${\rm F}_{\sigma}$&${\rm P}_{\sigma}$\\ 
States with $N_{\rm e}=1$&${\tilde {\cal R}}_{\sigma}$&${\rm A}_{\sigma}, {\rm B}_{\sigma}$&${\rm R}_{\sigma}$\\ 
States with $N_{\rm e}=3$&${\tilde {\cal A}}_{\sigma}$&${\rm F}_{\sigma}, {\rm G}_{\sigma}$&${\rm P}_{\sigma}$\\ 
Spin-triplet states&${\cal T}$&${\rm T}^{\gamma}$&${\rm T}^{\gamma}$\\\hline\hline 
\end{tabular}
\footnote[0]{$\sigma=\uparrow,\downarrow$ and $\gamma=1,0,-1$.}
\label{tbl:symbols}
\end{table}
\par
In the small-$|t|$ regime, effective states are obtained using the perturbation theory with respect to $t$. 
The ground state at half filling ($N_{\rm e}=2N_{\rm u}$) can be effectively expressed as 
\begin{equation}
|{\rm GS}\rangle^{2N_{\rm u}}=\prod_{i=1}^{N_{\rm u}}|{\cal G}\rangle_i,
\label{eq:GS}
\end{equation}
where $|{\cal G}\rangle_i$ denotes the ground state at unit cell $i$: 
$|{\cal G}\rangle_i=|\psi^-\rangle_i$ and $|{\rm S}\rangle_i$ for the Hubbard model and the KLM, respectively (Table \ref{tbl:symbols}). 
The ground-state energy is obtained as 
\begin{equation}
E_{\rm GS}^{2N_{\rm u}}=(E_{\cal G}+d\xi_{{\cal G}{\cal G}})N_{\rm u}
\end{equation}
in $d$ dimensions; $d=1$ for the ladder Hubbard model and 1D KLM, and $d=2$ for the bilayer Hubbard model and 2D KLM. 
Here, $\xi_{XY}$ denotes the bond energy of $\mathcal{O}(t^2)$ between states $X$ and $Y$ on neighboring unit cells, 
\begin{align}
\xi_{\psi^-\psi^-}&=-\frac{2t^2U^2}{(U^2+16t_{\perp}^2)^{\frac{3}{2}}},\quad\xi_{\psi^-{\rm T}}=-\frac{4t^2}{U}+\frac{2t^2}{\sqrt{U^2+16t_{\perp}^2}},\nonumber\\
\xi_{\psi^-{\rm A}}&=\xi_{\psi^-{\rm G}}=\frac{\xi_{\psi^-\psi^-}}{8}+\frac{3t^2}{8t_{\perp}}+\frac{3t^2}{2\sqrt{U^2+16t_{\perp}^2}},\nonumber\\
\xi_{\psi^-{\rm B}}&=\xi_{\psi^-{\rm F}}=\frac{\xi_{\psi^-\psi^-}}{8}-\frac{3t^2}{8t_{\perp}}+\frac{3t^2}{2\sqrt{U^2+16t_{\perp}^2}},\nonumber\\
\xi_{\psi^-{\rm V}}&=\xi_{\psi^-{\rm Z}}=\xi_{\psi^-{\rm D}^-}=\frac{4t^2}{U}+\frac{2t^2}{\sqrt{U^2+16t_{\perp}^2}}
\end{align}
in the Hubbard model \cite{KohnoHubLadder}, and 
\begin{equation}
\begin{array}{ccc}
\xi_{\rm SS}=-\frac{2t^2}{3J_{\rm K}},&\xi_{\rm ST}=-\frac{2t^2}{J_{\rm K}},&\xi_{\rm SP}=\xi_{\rm SR}=-\frac{3t^2}{4J_{\rm K}}
\end{array}
\end{equation}
in the KLM \cite{TsunetsuguRMP}. 
\par
The single-particle excited state with momentum ${\bm k}_{\parallel}$ for spin, charge, and electrons can be effectively expressed as 
\begin{equation}
|X({\bm k}_{\parallel})\rangle=\frac{1}{\sqrt{N_{\rm u}}}\sum_{i=1}^{N_{\rm u}}e^{i{\bm k}_{\parallel}\cdot{\bm r}_i}|X\rangle_i
\prod_{j(\ne i)}|{\cal G}\rangle_j,
\label{eq:Xk}
\end{equation}
where $|X\rangle$ denotes a unit-cell state in Tables \ref{tbl:Hub} and \ref{tbl:KLM}. 
For the spin excited states, $|X\rangle=|{\cal T}\rangle$, 
where $|{\cal T}\rangle$ represents the $N_{\rm e}=2$ spin-triplet ($S=1$) states in a unit cell: 
$|{\rm T}^1\rangle, |{\rm T}^{0}\rangle$, and $|{\rm T}^{-1}\rangle$ (Table \ref{tbl:symbols}). 
For the electron-addition excited states, $|X\rangle=|{\tilde {\cal A}}_{\sigma}\rangle$, 
where $|{\tilde {\cal A}}_{\sigma}\rangle$ represents the $N_{\rm e}=3$ states with $S=\frac{1}{2}$ in a unit cell: 
$|{\rm F}_{\sigma}\rangle$ and $|{\rm G}_{\sigma}\rangle$ in the Hubbard model and 
$|{\rm P}_{\sigma}\rangle$ in the KLM (Table \ref{tbl:symbols}). 
For the electron-removal excited states, $|X\rangle=|{\tilde {\cal R}}_{\sigma}\rangle$, 
where $|{\tilde {\cal R}}_{\sigma}\rangle$ represents the $N_{\rm e}=1$ states with $S=\frac{1}{2}$ in a unit cell: 
$|{\rm A}_{\sigma}\rangle$ and $|{\rm B}_{\sigma}\rangle$ in the Hubbard model and 
$|{\rm R}_{\sigma}\rangle$ in the KLM (Table \ref{tbl:symbols}). 
For the charge excited states, $|X\rangle=|{\rm D}^-\rangle$, $|{\rm V}\rangle$, and $|{\rm Z}\rangle$ in the Hubbard model. 
\par
The spin-excitation energy from the ground state is obtained using the second-order perturbation theory \cite{KohnoHubLadder,TsunetsuguRMP} as follows: 
\begin{equation}
e^{\rm spin}_{{\bm k}_{\parallel}}=J_{\rm eff}d\gamma_{{\bm k}_{\parallel}}+E_{\rm T}-E_{\cal G}+2d(\xi_{{\cal G}{\rm T}}-\xi_{{\cal G}{\cal G}}), 
\label{eq:spinExE}
\end{equation}
where $J_{\rm eff}$ denotes the effective exchange coupling of $\mathcal{O}(t^2)$ 
between neighboring $|{\cal T}\rangle$ and $|{\cal G}\rangle$, 
\begin{equation}
J_{\rm eff}=\left\{
\begin{array}{lllll}
\frac{8t^2}{U}-\frac{4t^2}{\sqrt{U^2+16t_{\perp}^2}}&{\rm for}&{\rm the}&{\rm Hubbard}&{\rm model},\\
\frac{4t^2}{J_{\rm K}}&{\rm for}&{\rm the}&{\rm KLM},&
\end{array}\right.
\end{equation}
and 
\begin{equation}
\label{eq:gammak}
\gamma_{{\bm k}_{\parallel}}=\frac{1}{d}\sum_{i=1}^d\cos k_{\parallel i} 
\end{equation}
with $k_{\parallel 1}=k_x$ and $k_{\parallel 2}=k_y$ in $d$ dimensions. 
\par
The electronic excitation energy is obtained as 
\begin{equation}
\varepsilon^X_{{\bm k}_{\parallel}}=-2t_{\rm eff}^Xd\gamma_{{\bm k}_{\parallel}}-2t_{\rm eff}^{3{\rm uc}}d(2d\gamma_{{\bm k}_{\parallel}}^2-1)+E_X-E_{\cal G}+2d(\xi_{{\cal G}X}-\xi_{{\cal G}{\cal G}}), 
\label{eq:eleExE}
\end{equation}
where $t_{\rm eff}^X$ denotes the effective inter-unit-cell hopping of $\mathcal{O}(t)$, 
\begin{equation}
t_{\rm eff}^{\rm A}=-t_{\rm eff}^-,\quad t_{\rm eff}^{\rm B}=-t_{\rm eff}^+,\quad t_{\rm eff}^{\rm F}=t_{\rm eff}^+,\quad t_{\rm eff}^{\rm G}=t_{\rm eff}^-
\label{eq:efftHubLadder}
\end{equation}
with $t_{\rm eff}^{\pm}=\frac{t}{2}\left(1\pm\frac{4t_\perp}{\sqrt{U^2+16t_{\perp}^2}}\right)$ for the Hubbard model \cite{KohnoHubLadder}, and 
\begin{equation}
\begin{array}{cc}
t_{\rm eff}^{\rm P}=\frac{t}{2},&t_{\rm eff}^{\rm R}=-\frac{t}{2}
\end{array}
\label{eq:efftKLM}
\end{equation}
for the KLM \cite{TsunetsuguRMP}. 
Here, $t_{\rm eff}^{3{\rm uc}}$ denotes the effective three-unit-cell hopping of $\mathcal{O}(t^2)$ \cite{TsunetsuguRMP}, 
\begin{equation}
t_{\rm eff}^{3{\rm uc}}=\left\{
\begin{array}{lllll}
-\frac{t^2U^2}{4(U^2+16t_{\perp}^2)^{\frac{3}{2}}}&{\rm for}&{\rm the}&{\rm Hubbard}&{\rm model},\\
-\frac{t^2}{6J_{\rm K}}&{\rm for}&{\rm the}&{\rm KLM}.&
\end{array}\right.
\end{equation}
\par
The charge-excitation energy from the ground state is obtained using the second-order perturbation theory as 
\begin{equation}
e^X_{{\bm k}_{\parallel}}=-2t_{\rm eff}^Xd\gamma_{{\bm k}_{\parallel}}+E_X-E_{\cal G}+2d(\xi_{{\cal G}X}-\xi_{{\cal G}{\cal G}}), 
\label{eq:chargeExE}
\end{equation}
where $t_{\rm eff}^X$ denotes the effective inter-unit-cell hopping of $\mathcal{O}(t^2)$, 
\begin{equation}
t_{\rm eff}^{{\rm D}^-}=-t_{\rm eff}^{\rm V}=-t_{\rm eff}^{\rm Z}=\xi_{\psi^-{\rm V}}.
\end{equation}
\par
\subsection{Excitation spectra at half filling at zero temperature} 
\label{sec:ekwT0}
\subsubsection{Electronic modes} 
At zero temperature, the dispersion relations of electronic modes at half filling are obtained from Eq. (\ref{eq:Akw}) as 
\begin{equation}
\begin{array}{lcl}
\omega=\varepsilon_{{\bm k}_{\parallel}}^X&{\rm for}&\langle X({\bm k}_{\parallel})|a_{{\bm k}_{\parallel},\sigma}^{\dagger}|{\rm GS}\rangle^{2N_{\rm u}}\ne 0,\\
\omega=-\varepsilon_{-{\bm k}_{\parallel}}^X&{\rm for}&\langle X(-{\bm k}_{\parallel})|a_{{\bm k}_{\parallel},\sigma}|{\rm GS}\rangle^{2N_{\rm u}}\ne 0, 
\end{array}
\label{eq:wXk}
\end{equation}
where $a_{{\bm k}_{\parallel},\sigma}^{\dagger}$ denotes the Fourier transform of $a_{i,\sigma}^{\dagger}$ for momentum ${\bm k}_{\parallel}$, 
\begin{equation}
a_{{\bm k}_{\parallel},\sigma}^{\dagger}=\frac{1}{\sqrt{N_{\rm u}}}\sum_{i=1}^{N_{\rm u}}e^{i{\bm k}_{\parallel}\cdot{\bm r}_i}a_{i,\sigma}^{\dagger}, 
\end{equation}
and $a_{i,\sigma}^{\dagger}$ denotes the creation operator of an electron with spin ${\sigma}$ at unit cell $i$, 
\begin{equation}
\begin{array}{lllll}
a_{i,\sigma}^{\dagger}=c_{i,\sigma}^{\dagger}&{\rm for}&{\rm the}&{\rm KLM},&\\
a_{i,\sigma}^{\dagger}=\frac{1}{\sqrt{2}}(c_{i_1,\sigma}^{\dagger}+e^{i k_{\perp}}c_{i_2,\sigma}^{\dagger})&{\rm for}&{\rm the}&{\rm Hubbard}&{\rm model}
\label{eq:aidagger}
\end{array}
\end{equation}
for sites $i_1$ and $i_2$ in unit cell $i$. 
\par
Specifically, in the Hubbard model \cite{KohnoHubLadder,localRungHubLadder}, 
\begin{equation}
\begin{array}{lclclcl}
\omega=\varepsilon_{{\bm k}_{\parallel}}^{\rm G}&{\rm and}&\omega=-\varepsilon_{-{\bm k}_{\parallel}}^{\rm B}&{\rm for}&k_{\perp}=0,\\
\omega=\varepsilon_{{\bm k}_{\parallel}}^{\rm F}&{\rm and}&\omega=-\varepsilon_{-{\bm k}_{\parallel}}^{\rm A}&{\rm for}&k_{\perp}=\pi 
\end{array}
\label{eq:wHub}
\end{equation}
because $|{\rm G}_{\sigma}\rangle$, $|{\rm B}_{\sigma}\rangle$, and $|{\rm GS}\rangle^{2N_{\rm u}}$ have $k_{\perp}=0$; 
$|{\rm F}_{\sigma}\rangle$ and $|{\rm A}_{\sigma}\rangle$ have $k_{\perp}=\pi$ [Table \ref{tbl:Hub}; Eqs. (\ref{eq:GS}) and (\ref{eq:Xk})]. 
In the KLM \cite {KohnoKLM,TsunetsuguRMP}, 
\begin{equation}
\begin{array}{lcl}
\omega=\varepsilon_{{\bm k}_{\parallel}}^{\rm P}&{\rm and}&\omega=-\varepsilon_{-{\bm k}_{\parallel}}^{\rm R}. 
\label{eq:wKLM}
\end{array}
\end{equation}
These modes [Eqs. (\ref{eq:wHub}) and (\ref{eq:wKLM}); solid brown curves 
in Figs. \ref{fig:HubLadder}(d-1)--\ref{fig:HubLadder}(f-1), \ref{fig:HubBilayer}(d-1)--\ref{fig:HubBilayer}(f-1), \ref{fig:KLM}(c-1), and \ref{fig:KLM}(d-1)] explain 
the dominant modes shown in Figs. \ref{fig:HubLadder}(a-1)--\ref{fig:HubLadder}(c-1), \ref{fig:HubBilayer}(a-1)--\ref{fig:HubBilayer}(c-1), \ref{fig:KLM}(a-1), and \ref{fig:KLM}(b-1). 
\subsubsection{Electronic continua} 
The two-particle excited state defined as 
\begin{align}
|XY({\bm k}_{\parallel};{\bm p}_{\parallel})\rangle&=\frac{1}{\sqrt{N_{\rm u}(N_{\rm u}-1)}}\nonumber\\
&\times\sum_{i\ne j}e^{i({\bm k}_{\parallel}-{\bm p}_{\parallel})\cdot{\bm r}_i}
e^{i{\bm p}_{\parallel}\cdot{\bm r}_j}|X\rangle_i|Y\rangle_j\prod_{l(\ne i,j)}|{\cal G}\rangle_l
\label{eq:XYk}
\end{align}
can also appear in the electronic spectrum, provided that 
\begin{equation}
\begin{array}{l}
\langle XY({\bm k}_{\parallel};{\bm p}_{\parallel})|a_{{\bm k}_{\parallel},\sigma}^{\dagger}|{\rm GS}\rangle^{2N_{\rm u}}\ne 0,\\
\langle XY(-{\bm k}_{\parallel};-{\bm p}_{\parallel})|a_{{\bm k}_{\parallel},\sigma}|{\rm GS}\rangle^{2N_{\rm u}}\ne 0.
\end{array}
\label{eq:condw2p}
\end{equation}
\par
In the ladder and bilayer Hubbard models, $|{\rm F_{\sigma^\prime}T^{\zeta}}({\bm k}_{\parallel};{\bm p}_{\parallel})\rangle$ and 
$|{\rm B_{{\bar \sigma}^\prime}T^{-\zeta}}(-{\bm k}_{\parallel};-{\bm p}_{\parallel})\rangle$, 
where $S^z$ of $\sigma$ is equal to that of $\sigma^\prime$ plus $\zeta$, can satisfy the condition [Eq. (\ref{eq:condw2p})] and form continua \cite{KohnoHubLadder}, 
\begin{equation}
\begin{array}{lcl}
\omega\approx\varepsilon_{{\bm k}_{\parallel}-{\bm p}_{\parallel}}^{\rm F}+e^{\rm spin}_{{\bm p}_{\parallel}}&{\rm for}&k_{\perp}=0,\\
\omega\approx-\varepsilon_{-{\bm k}_{\parallel}+{\bm p}_{\parallel}}^{\rm B}-e^{\rm spin}_{-{\bm p}_{\parallel}}&{\rm for}&k_{\perp}=\pi, 
\end{array}
\label{eq:w2pHub}
\end{equation}
respectively [light-yellow regions in Figs. \ref{fig:HubLadder}(d-1)--\ref{fig:HubLadder}(f-1) and \ref{fig:HubBilayer}(d-1)--\ref{fig:HubBilayer}(f-1)]. 
The corresponding continua can be seen in the electronic spectra 
in Figs. \ref{fig:HubLadder}(a-1)--\ref{fig:HubLadder}(c-1) and \ref{fig:HubBilayer}(a-1)--\ref{fig:HubBilayer}(c-1). 
\par
Similarly, in the 1D and 2D KLMs, $|{\rm P_{\sigma^\prime}T^{\zeta}}({\bm k}_{\parallel};{\bm p}_{\parallel})\rangle$ and 
$|{\rm R_{{\bar \sigma}^\prime}T^{-\zeta}}(-{\bm k}_{\parallel};-{\bm p}_{\parallel})\rangle$, 
where $S^z$ of $\sigma$ is equal to that of $\sigma^\prime$ plus $\zeta$, can satisfy the condition [Eq. (\ref{eq:condw2p})] and form continua, 
\begin{equation}
\begin{array}{ll}
\omega&\approx\varepsilon_{{\bm k}_{\parallel}-{\bm p}_{\parallel}}^{\rm P}+e^{\rm spin}_{{\bm p}_{\parallel}},\\
\omega&\approx-\varepsilon_{-{\bm k}_{\parallel}+{\bm p}_{\parallel}}^{\rm R}-e^{\rm spin}_{-{\bm p}_{\parallel}}, 
\end{array}
\label{eq:w2pKLM}
\end{equation}
respectively [light-yellow regions in Figs. \ref{fig:KLM}(c-1) and \ref{fig:KLM}(d-1)]. 
The corresponding continua can be seen in the electronic spectra in Figs. \ref{fig:KLM}(a-1) and \ref{fig:KLM}(b-1). 
\subsubsection{Spin mode and particle-hole continuum} 
\begin{figure*}
\includegraphics[width=\linewidth]{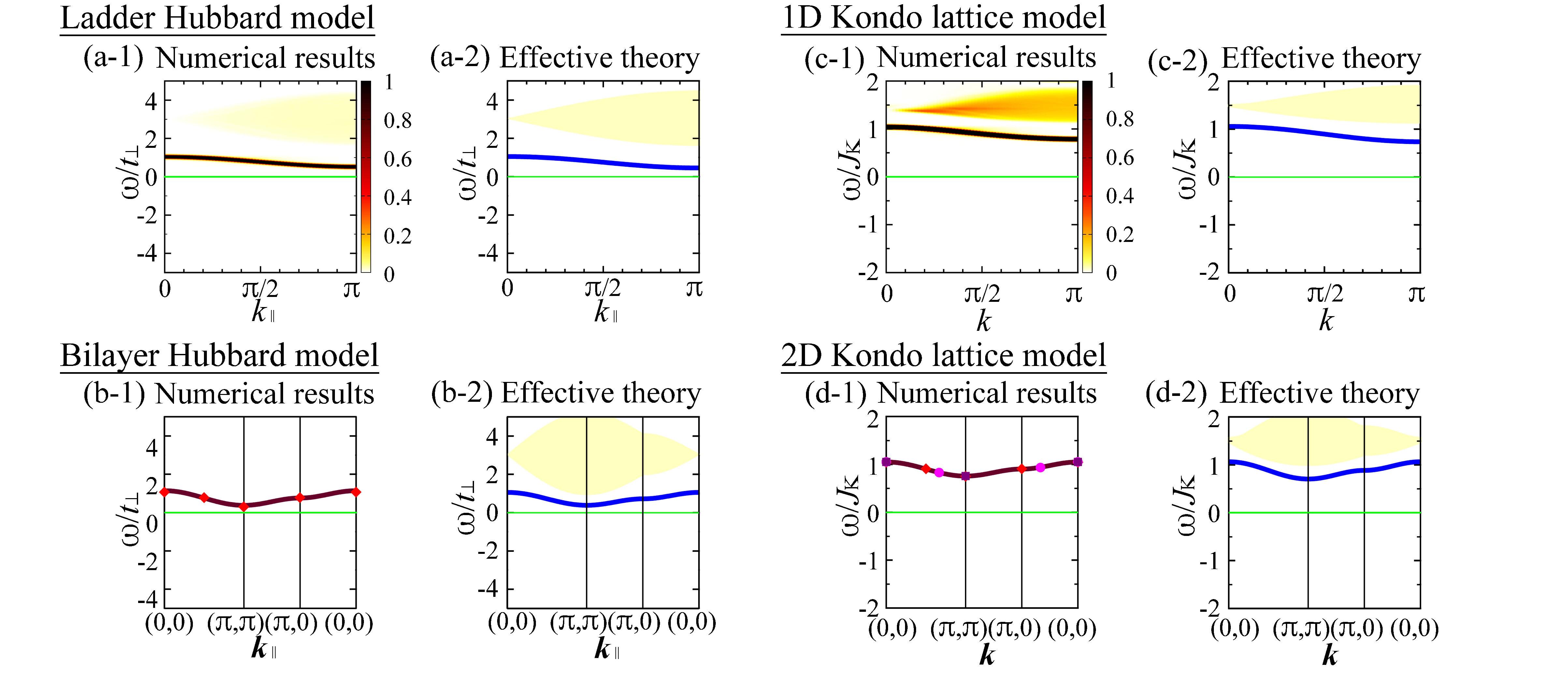}
\caption{Spin excitation of the ladder Hubbard model for $t/t_{\perp}=0.4$ and $U/t_{\perp}=3$ [(a-1), (a-2)], 
the bilayer Hubbard model for $t/t_{\perp}=0.3$ and $U/t_{\perp}=3$ [(b-1), (b-2)], 
the 1D Kondo lattice model for $t/J_{\rm K}=0.2$ [(c-1), (c-2)], and the 2D Kondo lattice model for $t/J_{\rm K}=0.15$ [(d-1), (d-2)] at $T=0$ and $\delta=0$. 
(a-1) $S_{\rm Hub}(k_{\parallel},\omega)t_{\perp}/3$, (c-1) $S_{\rm KLM}(k,\omega)J_{\rm K}/3$ obtained using the non-Abelian DDMRG method. 
Gaussian broadening is used with a standard deviation of $0.05t_{\perp}$ [(a-1)] and $0.02J_{\rm K}$ [(c-1)]. 
(b-1), (d-1) The lowest spin-excitation energy at each momentum ${\bm k}_{\parallel}$ (for $k_{\perp}=\pi$ in the bilayer Hubbard model) 
obtained using the Lanczos method on lattices of $\sqrt{10}\times\sqrt{10}$ unit cells (purple squares) [(d-1)], $3\times 3$ unit cells (magenta circles) [(d-1)], and $\sqrt{8}\times\sqrt{8}$ unit cells (red diamonds) [(b-1), (d-1)]. 
The sold brown curves indicate the least-squares fitting in the form of $\omega=Jd\gamma_{{\bm k}_{\parallel}}+\Delta E$ using the exact-diagonalization data [(b-1), (d-1)]. 
(a-2)--(d-2) Dispersion relations obtained using the effective theory for weak inter-unit-cell hopping. 
The solid blue curves represent $\omega=e^{\rm spin}_{{\bm k}_{\parallel}}$ [(a-2)--(d-2)]. 
The light-yellow regions indicate particle-hole continua: $\omega=\varepsilon^{\rm F}_{{\bm k}_{\parallel}-{\bm p}_{\parallel}}+\varepsilon^{\rm B}_{{\bm p}_{\parallel}}$ [(a-2), (b-2)] 
and $\omega=\varepsilon^{\rm P}_{{\bm k}-{\bm p}}+\varepsilon^{\rm R}_{\bm p}$ [(c-2), (d-2)]. 
The solid green lines indicate $\omega=0$.}
\label{fig:spin}
\end{figure*}

The dispersion relation of the spin mode is expressed as 
\begin{equation}
\omega=e_{{\bm k}_{\parallel}}^{\rm spin} 
\label{eq:wspin}
\end{equation}
[Eq. (\ref{eq:spinExE})] \cite{KohnoHubLadder,KohnoKLM,TsunetsuguRMP}, 
which explains the numerical results reasonably well [Figs. \ref{fig:spin}(a-1)--\ref{fig:spin}(d-1) and \ref{fig:spin}(a-2)--\ref{fig:spin}(d-2)]. 
The particle-hole continuum of $|{\rm F}_{\sigma}{\rm B}_{\sigma}({\bm k}_{\parallel};{\bm p}_{\parallel})\rangle$ and 
$\frac{1}{\sqrt{2}}\sum_{\sigma}|{\rm F}_{\sigma}{\rm B}_{\bar \sigma}({\bm k}_{\parallel};{\bm p}_{\parallel})\rangle$ in the ladder and bilayer Hubbard models 
and that of $|{\rm P}_{\sigma}{\rm R}_{\sigma}({\bm k}_{\parallel};{\bm p}_{\parallel})\rangle$ and 
$\frac{1}{\sqrt{2}}\sum_{\sigma}|{\rm P}_{\sigma}{\rm R}_{\bar \sigma}({\bm k}_{\parallel};{\bm p}_{\parallel})\rangle$ in the 1D and 2D KLMs \cite{KohnoKLM}, 
\begin{equation}
\begin{array}{l}
\omega\approx\varepsilon_{{\bm k}_{\parallel}-{\bm p}_{\parallel}}^{\rm F}+\varepsilon_{{\bm p}_{\parallel}}^{\rm B},\\
\omega\approx\varepsilon_{{\bm k}_{\parallel}-{\bm p}_{\parallel}}^{\rm P}+\varepsilon_{{\bm p}_{\parallel}}^{\rm R},
\end{array}
\label{eq:phcont}
\end{equation}
respectively [light-yellow regions in Figs. \ref{fig:spin}(a-2)--\ref{fig:spin}(d-2)], appear above the spin modes [Figs. \ref{fig:spin}(a-1) and \ref{fig:spin}(c-1)]. 
\subsection{Doping-induced states at zero temperature} 
\label{sec:DIS}
\subsubsection{Hole-doping-induced states} 
\label{sec:holeDIS}
The one-hole-doped ground state, which is the lowest-energy state in the subspace of $N_{\rm e}=2N_{\rm u}-1$, is 
$|{\cal R}_{\sigma}(-{\bm k}_{\parallel{\rm F}}^-)\rangle$ with ${\bm k}_{\parallel{\rm F}}^-={\bm \pi}$ for $t>0$ 
[Eqs. (\ref{eq:eleExE}), (\ref{eq:efftHubLadder}), and (\ref{eq:efftKLM})] \cite{KohnoHubLadder,KohnoKLM}, 
where $|{\cal R}_{\sigma}\rangle$ represents the $N_{\rm e}=1$ low-energy state in a unit cell: 
$|{\rm B}_{\sigma}\rangle$ for the Hubbard model and $|{\rm R}_{\sigma}\rangle$ for the KLM (Tables \ref{tbl:Hub}, \ref{tbl:KLM}, and \ref{tbl:symbols}). 
Here, ${\bm k}_{\parallel{\rm F}}^-$ denotes the Fermi momentum in the small-hole-doping limit, 
i.e., the momentum ${\bm k}_{\parallel}$ at the top of the lower band at zero temperature. 
The energy of the one-hole-doped ground state is obtained as 
\begin{equation}
E_{\rm GS}^{2N_{\rm u}-1}=\varepsilon^{\cal R}_{-{\bm k}_{\parallel{\rm F}}^-}+E_{\rm GS}^{2N_{\rm u}}. 
\label{eq:1holeEGS}
\end{equation}
The chemical potential $\mu$ is adjusted such that 
\begin{equation}
E_{\rm GS}^{2N_{\rm u}-1}=E_{\rm GS}^{2N_{\rm u}}
\label{eq:1holeEGS2}
\end{equation}
for the one-hole-doped system at zero temperature. 
\par
The electron-addition excited state in the one-hole-doped system at zero temperature is obtained as 
\begin{align}
&a_{{\bm k}_{\parallel},\sigma}^{\dagger}|{\rm GS}\rangle^{2N_{\rm u}-1}_{\sigma^{\prime}}
=\frac{1}{N_{\rm u}}\sum_{i,j}e^{i{\bm k}_{\parallel}\cdot{\bm r}_i}e^{-i{\bm k}_{\parallel{\rm F}}^-\cdot{\bm r}_j}
a_{i,\sigma}^{\dagger}|{\cal R}_{\sigma^{\prime}}\rangle_j\prod_{l(\ne j)}|{\cal G}\rangle_l\\
&=\frac{1}{N_{\rm u}}\sum_{i\ne j}e^{i{\bm k}_{\parallel}\cdot{\bm r}_i}e^{-i{\bm k}_{\parallel{\rm F}}^-\cdot{\bm r}_j}
a_{i,\sigma}^{\dagger}|{\cal G}\rangle_i|{\cal R}_{\sigma^{\prime}}\rangle_j\prod_{l(\ne i,j)}|{\cal G}\rangle_l\nonumber\\
&+\frac{1}{N_{\rm u}}\sum_{i}e^{i({\bm k}_{\parallel}-{\bm k}_{\parallel{\rm F}}^-)\cdot{\bm r}_i}
a_{i,\sigma}^{\dagger}|{\cal R}_{\sigma^{\prime}}\rangle_i\prod_{l(\ne i)}|{\cal G}\rangle_l,
\label{eq:add}
\end{align}
where $|{\rm GS}\rangle^{2N_{\rm u}-1}_{\sigma^{\prime}}$ represents the one-hole-doped ground state with spin $\sigma^{\prime}$. 
\par
The energy of the first term of Eq. (\ref{eq:add}) is approximately 
$\varepsilon^{\tilde {\cal A}}_{{\bm k}_{\parallel}}+\varepsilon^{\cal R}_{-{\bm k}_{\parallel{\rm F}}^-}+E_{\rm GS}^{2N_{\rm u}}$, 
where $|{\tilde {\cal A}}_{\sigma}\rangle$ represents the $N_{\rm e}=3$ states in a unit cell (Table \ref{tbl:symbols}). 
Because $E_{\rm GS}^{2N_{\rm u}-1}=\varepsilon^{\cal R}_{-{\bm k}_{\parallel{\rm F}}^-}+E_{\rm GS}^{2N_{\rm u}}$ [Eq. (\ref{eq:1holeEGS})], 
the states of the first term of Eq. (\ref{eq:add}) appear in the electron-addition spectrum of the one-hole-doped system, exhibiting the dispersion relations of 
$\omega\approx\varepsilon^{\tilde {\cal A}}_{{\bm k}_{\parallel}}$, which are basically the same as those at half filling [Eqs. (\ref{eq:wHub}) and (\ref{eq:wKLM})]. 
\par
When ${\bm k}_{\parallel}={\bm k}_{\parallel{\rm F}}^-$ and $\sigma^{\prime}={\bar \sigma}$, 
the second term of Eq. (\ref{eq:add}) includes the component of $|{\rm GS}\rangle^{2N_{\rm u}}$ 
[Eq. (\ref{eq:GS}); ${}_i\langle{\cal G}|a_{i,\sigma}^{\dagger}|{\cal R}_{\bar \sigma}\rangle_i\ne 0$ (for $k_{\perp}=0$ in the Hubbard model) 
(Tables \ref{tbl:Hub}, \ref{tbl:KLM}, and \ref{tbl:symbols})] 
with energy $E_{\rm GS}^{2N_{\rm u}}(=E_{\rm GS}^{2N_{\rm u}-1})$ [Eq. (\ref{eq:1holeEGS2})], which corresponds to the top of the lower band. 
\par
The second term of Eq. (\ref{eq:add}) also includes the component of $|{\cal T}({\bm k}_{\parallel}-{\bm k}_{\parallel{\rm F}}^-)\rangle$ 
[Eq. (\ref{eq:Xk}); ${}_i\langle{\cal T}|a_{i,\sigma}^{\dagger}|{\cal R}_{\sigma^{\prime}}\rangle_i\ne 0$ (for $k_{\perp}=\pi$ in the Hubbard model) 
(Tables \ref{tbl:Hub}, \ref{tbl:KLM}, and \ref{tbl:symbols})] 
with energy $e^{\rm spin}_{{\bm k}_{\parallel}-{\bm k}_{\parallel{\rm F}}^-}+E_{\rm GS}^{2N_{\rm u}}$. 
This component appears in the electron-addition spectrum (for $k_{\perp}=\pi$ in the Hubbard model) 
in the one-hole-doped system \cite{KohnoHubLadder,KohnoKLM,KohnoDIS} along 
\begin{equation}
\label{eq:ekHoleDIS}
\omega=e^{\rm spin}_{{\bm k}_{\parallel}-{\bm k}_{\parallel{\rm F}}^-}+E_{\rm GS}^{2N_{\rm u}}-E_{\rm GS}^{2N_{\rm u}-1}
\overset{{\rm Eq.} (\ref{eq:1holeEGS2})}{=}e^{\rm spin}_{{\bm k}_{\parallel}-{\bm k}_{\parallel{\rm F}}^-}. 
\end{equation}
Thus, the electronic mode emerges along $\omega=e^{\rm spin}_{{\bm k}_{\parallel}-{\bm k}_{\parallel{\rm F}}^-}$ in the one-hole-doped system, 
as indicated by the solid red curves in Figs. \ref{fig:HubLadder}(d-2), \ref{fig:HubLadder}(f-2), \ref{fig:HubBilayer}(d-2), \ref{fig:HubBilayer}(f-2), 
\ref{fig:KLM}(c-2), and \ref{fig:KLM}(d-2) 
[Figs. \ref{fig:HubLadder}(a-2), \ref{fig:HubLadder}(c-2), \ref{fig:HubBilayer}(a-2), \ref{fig:HubBilayer}(c-2), \ref{fig:KLM}(a-2), and \ref{fig:KLM}(b-2)]. 
In general, by doping Mott and Kondo insulators, spin excited states emerge in the electronic spectrum, 
exhibiting the magnetic dispersion relation shifted by the Fermi momentum 
\cite{Kohno1DHub,Kohno2DHub,KohnoDIS,KohnoRPP,KohnoHubLadder,KohnoKLM,KohnoGW,KohnoAF,KohnoSpin,Kohno1DtJ,Kohno2DtJ}. 
\par
\begin{figure*}
\includegraphics[width=\linewidth]{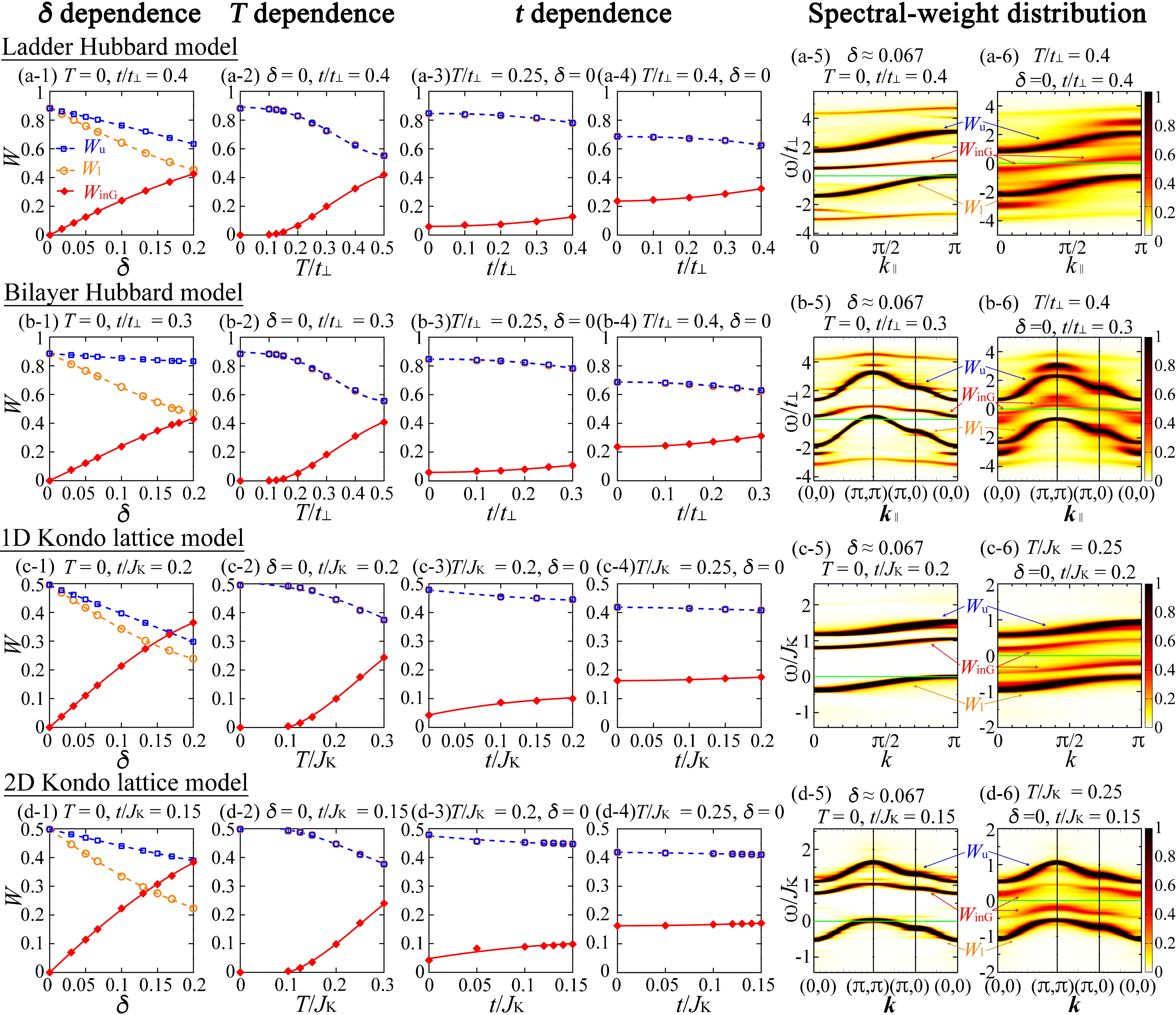}
\caption{Doping dependence, temperature dependence, and inter-unit-cell-hopping dependence of spectral weights. 
(a-1)--(d-1) $\delta$ dependence at $T=0$, (a-2)--(d-2) $T$ dependence at $\delta=0$ 
in the ladder Hubbard model for $t/t_{\perp}=0.4$ and $U/t_{\perp}=3$ [(a-1), (a-2)]; 
the bilayer Hubbard model for $t/t_{\perp}=0.3$ and $U/t_{\perp}=3$ [(b-1), (b-2)]; 
the 1D KLM for $t/J_{\rm K}=0.2$ [(c-1), (c-2)]; the 2D KLM for $t/J_{\rm K}=0.15$ [(d-1), (d-2)]. 
(a-3)--(d-3), (a-4)--(d-4) $t$ dependence at $\delta=0$ in the ladder Hubbard model for $U/t_{\perp}=3$ at $T/t_{\perp}=0.25$ [(a-3)] and 0.4 [(a-4)]; 
the bilayer Hubbard model for $U/t_{\perp}=3$ at $T/t_{\perp}=0.25$ [(b-3)] and 0.4 [(b-4)]; 
the 1D KLM at $T/J_{\rm K}=0.2$ [(c-3)] and 0.25 [(c-4)]; the 2D KLM at $T/J_{\rm K}=0.2$ [(d-3)] and 0.25 [(d-4)]. 
In (a-1)--(a-4), (b-1)--(b-4), (c-1)--(c-4), and (d-1)--(d-4), the spectral weights of the upper dominant band $W_{\rm u}$ (open blue squares with a dashed blue curve), 
the lower dominant band $W_{\rm l}$ (open orange circles with a dashed orange curve), and the in-gap bands $W_{\rm inG}$ (solid red diamonds with a solid red curve) 
indicated in (a-5)--(d-5) and (a-6)--(d-6) are shown. 
The curves are guides for eyes. 
(a-5), (a-6) $A(k_{\parallel},0,\omega)t_{\perp}+A(k_{\parallel},\pi,\omega)t_{\perp}$ of the ladder Hubbard model for $t/t_{\perp}=0.4$ and $U/t_{\perp}=3$, 
(b-5), (b-6) ${\bar A}({\bm k}_{\parallel},0,\omega)t_{\perp}+{\bar A}({\bm k}_{\parallel},\pi,\omega)t_{\perp}$ 
of the bilayer Hubbard model for $t/t_{\perp}=0.3$ and $U/t_{\perp}=3$, 
(c-5), (c-6) $A(k,\omega)J_{\rm K}$ of the 1D KLM for $t/J_{\rm K}=0.2$, 
(d-5), (d-6) ${\bar A}({\bm k},\omega)J_{\rm K}$ for $t/J_{\rm K}=0.15$ 
at $\delta\approx 0.067$ and $T=0$ [(a-5)--(d-5)]; $\delta=0$ and $T/t_{\perp}=0.4$ [(a-6), (b-6)]; $\delta=0$ and $T/J_{\rm K}=0.25$ [(c-6), (d-6)]. 
In (a-5)--(d-5) and (a-6)--(d-6), the upper dominant band, the in-gap bands, and the lower dominant band are indicated 
by $W_{\rm u}$, $W_{\rm inG}$, and $W_{\rm l}$, respectively. 
Gaussian broadening is used with a standard deviation of $0.05t_{\perp}$ [(a-5), (a-6), (b-5), (b-6)] and $0.02J_{\rm K}$ [(c-5), (c-6), (d-5), (d-6)]. 
The solid green lines indicate $\omega=0$.}
\label{fig:wt}
\end{figure*}

Because the component of $|{\cal T}({\bm k}_{\parallel}-{\bm k}_{\parallel{\rm F}}^-)\rangle$ in the second term of Eq. (\ref{eq:add}) is 
$\frac{1}{\sqrt{N_{\rm u}}}|{\cal T}({\bm k}_{\parallel}-{\bm k}_{\parallel{\rm F}}^-)\rangle W_{\cal TR}$, 
where $W_{\cal TR}={}_i\langle{\cal T}|a_{i,\sigma}^{\dagger}|{\cal R}_{\sigma^{\prime}}\rangle_i=\mathcal{O}(1)$, 
the spectral weight is $\mathcal{O}(\frac{1}{N_{\rm u}})$ in the one-hole-doped system. 
In the $N_{\rm h}$-hole-doped system where $N_{\rm h}$ unit cells are $|{\cal R}_{\sigma^{\prime}}\rangle$, 
the probability that an electron is added to $|{\cal R}_{\sigma^{\prime}}\rangle$ becomes $N_{\rm h}$ times as large as that of the one-hole-doped system, and 
the spectral weight of the emergent mode becomes $\mathcal{O}(\frac{N_{\rm h}}{N_{\rm u}})$. 
This implies that the spectral weight of the doping-induced states is proportional to the doping concentration in the small-doping regime [Figs. \ref{fig:wt}(a-1)--\ref{fig:wt}(d-1)] 
\cite{Kohno1DHub,Kohno2DHub,KohnoDIS,KohnoRPP,KohnoHubLadder,KohnoKLM,KohnoGW,KohnoAF,KohnoSpin,Kohno1DtJ,Kohno2DtJ,Eskes,DagottoDOS}. 
\par
The dispersion relation of the doping-induced mode remains almost unchanged in the small-doping regime. 
This can be explained as follows: 
In the $N_{\rm h}$-hole-doped system, the spin excited states in the $(N_{\rm h}-1)$-hole-doped system appears as electron-addition excited states. 
The ground-state momentum in the $N_{\rm h}$-hole-doped system always differs from that of the $(N_{\rm h}-1)$-hole-doped system by the Fermi momentum. 
Hence, the spin excited states of the $(N_{\rm h}-1)$-hole-doped system appears in the electronic spectrum of the $N_{\rm h}$-hole-doped system 
with the magnetic dispersion relation shifted by the Fermi momentum. 
In the small-doping regime, the Fermi momentum and the spin-excitation dispersion relation are almost the same as those in the small-doping limit. 
\par
In the ladder and bilayer Hubbard models, the second term of Eq. (\ref{eq:add}) also includes 
the component of $|{\rm D}^-({\bm k}_{\parallel}-{\bm k}_{\parallel{\rm F}}^-)\rangle$ 
because ${}_i\langle{\rm D}^-|a_{i,\sigma}^{\dagger}|{\rm B}_{{\bar \sigma}}\rangle_i\ne 0$ for $k_{\perp}=\pi$. 
The dispersion relation of the corresponding doping-induced mode is obtained as 
\begin{equation}
\label{eq:ekD-}
\omega=e^{{\rm D}^-}_{{\bm k}_{\parallel}-{\bm k}_{\parallel{\rm F}}^-}
\end{equation}
[dotted red curves in Figs. \ref{fig:HubLadder}(d-2), \ref{fig:HubLadder}(f-2), \ref{fig:HubBilayer}(d-2), and \ref{fig:HubBilayer}(f-2)]. 
In addition, because ${}_i\langle{\rm V}|a_{i,\sigma}|{\rm B}_{\sigma}\rangle_i\ne 0$ for $k_{\perp}=0$, 
the doping-induced mode appears for $\omega<0$, exhibiting 
\begin{equation}
\label{eq:ekV}
\omega=-e^{\rm V}_{-{\bm k}_{\parallel}-{\bm k}_{\parallel{\rm F}}^-}
\end{equation}
[dotted magenta curves in Figs. \ref{fig:HubLadder}(d-2), \ref{fig:HubLadder}(e-2), \ref{fig:HubBilayer}(d-2), and \ref{fig:HubBilayer}(e-2)]. 
These high-$|\omega|$ modes can be seen in the numerical results [Figs. \ref{fig:HubLadder}(a-2)--\ref{fig:HubLadder}(c-2) and \ref{fig:HubBilayer}(a-2)--\ref{fig:HubBilayer}(c-2)]. 
The high-$|\omega|$ doping-induced modes have also been shown to appear in the periodic Anderson model \cite{KohnoKLM}. 
\subsubsection{Electron-doping-induced states} 
\label{sec:eleDIS}
The one-electron-doped ground state, which is the lowest-energy state in the subspace of $N_{\rm e}=2N_{\rm u}+1$, is 
$|{\cal A}_{\sigma}({\bm k}_{\parallel{\rm F}}^+)\rangle$ with ${\bm k}_{\parallel{\rm F}}^+={\bm 0}$ for $t>0$ 
[Eqs. (\ref{eq:eleExE}), (\ref{eq:efftHubLadder}), and (\ref{eq:efftKLM})], 
where $|{\cal A}_{\sigma}\rangle$ represents the $N_{\rm e}=3$ low-energy state in a unit cell: 
$|{\rm F}_{\sigma}\rangle$ for the Hubbard model and $|{\rm P}_{\sigma}\rangle$ for the KLM (Tables \ref{tbl:Hub}, \ref{tbl:KLM}, and \ref{tbl:symbols}). 
Here, ${\bm k}_{\parallel{\rm F}}^+$ denotes the Fermi momentum in the small-electron-doping limit, 
i.e., the momentum ${\bm k}_{\parallel}$ at the bottom of the upper band at zero temperature. 
The energy of the one-electron-doped ground state is obtained as 
\begin{equation}
E_{\rm GS}^{2N_{\rm u}+1}=\varepsilon^{\cal A}_{{\bm k}_{\parallel{\rm F}}^+}+E_{\rm GS}^{2N_{\rm u}}. 
\label{eq:1eleEGS}
\end{equation}
The chemical potential $\mu$ is adjusted such that 
\begin{equation}
E_{\rm GS}^{2N_{\rm u}+1}=E_{\rm GS}^{2N_{\rm u}}
\label{eq:1eleEGS2}
\end{equation}
for the one-electron-doped system at zero temperature. 
\par
The electron-removal excited state in the one-electron-doped system at zero temperature is obtained as 
\begin{align}
&a_{{\bm k}_{\parallel},\sigma}|{\rm GS}\rangle^{2N_{\rm u}+1}_{\sigma^{\prime}}
=\frac{1}{N_{\rm u}}\sum_{i,j}e^{-i{\bm k}_{\parallel}\cdot{\bm r}_i}e^{i{\bm k}_{\parallel{\rm F}}^+\cdot{\bm r}_j}
a_{i,\sigma}|{\cal A}_{\sigma^{\prime}}\rangle_j\prod_{l(\ne j)}|{\cal G}\rangle_l\\
&=\frac{1}{N_{\rm u}}\sum_{i\ne j}e^{-i{\bm k}_{\parallel}\cdot{\bm r}_i}e^{i{\bm k}_{\parallel{\rm F}}^+\cdot{\bm r}_j}
a_{i,\sigma}|{\cal G}\rangle_i|{\cal A}_{\sigma^{\prime}}\rangle_j\prod_{l(\ne i,j)}|{\cal G}\rangle_l\nonumber\\
&+\frac{1}{N_{\rm u}}\sum_{i}e^{i(-{\bm k}_{\parallel}+{\bm k}_{\parallel{\rm F}}^+)\cdot{\bm r}_i}
a_{i,\sigma}|{\cal A}_{\sigma^{\prime}}\rangle_i\prod_{l(\ne i)}|{\cal G}\rangle_l,
\label{eq:rmv}
\end{align}
where $|{\rm GS}\rangle^{2N_{\rm u}+1}_{\sigma^{\prime}}$ represents the one-electron-doped ground state with spin $\sigma^{\prime}$. 
\par
The states described in the first term of Eq. (\ref{eq:rmv}) form the modes exhibiting essentially the same dispersion relations as those at half filling: 
$\omega\approx -\varepsilon^{\tilde {\cal R}}_{-{\bm k}_{\parallel}}$, because the energy is approximately 
$\varepsilon^{\tilde {\cal R}}_{-{\bm k}_{\parallel}}+\varepsilon^{\cal A}_{{\bm k}_{\parallel{\rm F}}^+}+E_{\rm GS}^{2N_{\rm u}}
=\varepsilon^{\tilde {\cal R}}_{-{\bm k}_{\parallel}}+E_{\rm GS}^{2N_{\rm u}+1}$ [Eq. (\ref{eq:1eleEGS})], 
as in the hole-doped case [Sec. \ref{sec:holeDIS}; Eqs. (\ref{eq:wHub}) and (\ref{eq:wKLM})]. 
Here, $|{\tilde {\cal R}}_{\bar {\sigma}}\rangle$ represents the $N_{\rm e}=1$ states in a unit cell (Table \ref{tbl:symbols}). 
\par
When ${\bm k}_{\parallel}={\bm k}_{\parallel{\rm F}}^+$ and $\sigma^{\prime}=\sigma$, 
the second term of Eq. (\ref{eq:rmv}) includes the component of $|{\rm GS}\rangle^{2N_{\rm u}}$ 
[Eq. (\ref{eq:GS}); ${}_i\langle{\cal G}|a_{i,\sigma}|{\cal A}_{\sigma}\rangle_i\ne 0$ (for $k_{\perp}=\pi$ in the Hubbard model) 
(Tables \ref{tbl:Hub}, \ref{tbl:KLM}, and \ref{tbl:symbols})] 
with energy $E_{\rm GS}^{2N_{\rm u}}(=E_{\rm GS}^{2N_{\rm u}+1})$ [Eq. (\ref{eq:1eleEGS2})], which corresponds to the bottom of the upper band. 
\par
The second term of Eq. (\ref{eq:rmv}) also includes the component of $|{\cal T}(-{\bm k}_{\parallel}+{\bm k}_{\parallel{\rm F}}^+)\rangle$ 
[Eq. (\ref{eq:Xk}); ${}_i\langle{\cal T}|a_{i,\sigma}|{\cal A}_{\sigma^{\prime}}\rangle_i\ne 0$ (for $k_{\perp}=0$ in the Hubbard model) 
(Tables \ref{tbl:Hub}, \ref{tbl:KLM}, and \ref{tbl:symbols})] 
with energy $e^{\rm spin}_{-{\bm k}_{\parallel}+{\bm k}_{\parallel{\rm F}}^+}+E_{\rm GS}^{2N_{\rm u}}$. 
This component appears in the electron-removal spectrum (for $k_{\perp}=0$ in the Hubbard model) in the one-electron-doped system along 
\begin{equation}
\label{eq:ekEleDIS}
\omega=-e^{\rm spin}_{-{\bm k}_{\parallel}+{\bm k}_{\parallel{\rm F}}^+}-E_{\rm GS}^{2N_{\rm u}}+E_{\rm GS}^{2N_{\rm u}+1}
\overset{{\rm Eq.} (\ref{eq:1eleEGS2})}{=}-e^{\rm spin}_{-{\bm k}_{\parallel}+{\bm k}_{\parallel{\rm F}}^+}. 
\end{equation}
Thus, the electronic mode emerges along $\omega=-e^{\rm spin}_{-{\bm k}_{\parallel}+{\bm k}_{\parallel{\rm F}}^+}$ in the one-electron-doped system 
\cite{Kohno1DHub,Kohno2DHub,KohnoDIS,KohnoRPP,KohnoHubLadder,KohnoKLM,KohnoGW,KohnoAF,KohnoSpin,Kohno1DtJ,Kohno2DtJ}. 
\par
The high-$|\omega|$ modes also appear in the ladder and bilayer Hubbard models, exhibiting 
\begin{equation}
\label{eq:ekZ}
\begin{array}{lll}
\omega=e^{\rm Z}_{{\bm k}_{\parallel}+{\bm k}_{\parallel{\rm F}}^+}&{\rm for}&k_{\perp}=\pi,\\
\omega=-e^{{\rm D}^-}_{-{\bm k}_{\parallel}+{\bm k}_{\parallel{\rm F}}^+}&{\rm for}&k_{\perp}=0,
\end{array}
\end{equation}
because ${}_i\langle{\rm Z}|a_{i,\sigma}^{\dagger}|{\rm F}_{\bar \sigma}\rangle_i\ne 0$ for $k_{\perp}=\pi$ and 
${}_i\langle{\rm D}^-|a_{i,\sigma}|{\rm F}_{\sigma}\rangle_i\ne 0$ for $k_{\perp}=0$. 
\par
The spectral weights of the emergent modes increase in proportion to the doping concentration $|\delta|$ 
with the dispersion relations almost unchanged in the small-doping regime, as in the hole-doped case (Sec. \ref{sec:holeDIS}). 
The features of electron-doped systems can also be explained based on the results for hole-doped systems (Sec. \ref{sec:holeDIS}) 
using the particle-hole and gauge transformations on bipartite lattices \cite{Essler,TakahashiBook}. 
\subsection{Temperature-induced states} 
\label{sec:TindStates}
\subsubsection{Dispersion relations of temperature-induced modes} 
\label{sec:TindModes}
At nonzero temperature, the thermal state [Eq. (\ref{eq:thermalState})] involves all the eigenstates as follows: 
\begin{align}
\label{eq:thermalState2}
|\beta\rangle_l&=e^{-\frac{\beta{\cal H}}{2}}|{\rm I}\rangle_l=\sum_n e^{-\frac{\beta E_n}{2}}|n\rangle\langle n|{\rm I}\rangle_l\nonumber\\
&=e^{-\frac{\beta E_{\rm GS}^{2N_{\rm u}}}{2}}\sum_n e^{-\frac{\beta \Delta E_n}{2}}|n\rangle\langle n|{\rm I}\rangle_l, 
\end{align}
where $\Delta E_n=E_n-E_{\rm GS}^{2N_{\rm u}}$. 
At zero temperature, the thermal state consists only of the ground state. 
At low temperatures, low-energy excited states ($\beta\Delta E_n\lesssim 1$) can also significantly contribute to the thermal state. 
If $\beta(E_{\rm GS}^{2N_{\rm u}\pm 1}-E_{\rm GS}^{2N_{\rm u}})\lesssim 1$, the one-hole-doped ground state $|{\rm GS}\rangle_{\sigma}^{2N_{\rm u}-1}$ 
and one-electron-doped ground state $|{\rm GS}\rangle_{\sigma}^{2N_{\rm u}+1}$ can significantly contribute to the thermal state. 
In this case, electronic states emerge along 
\begin{align}
\label{eq:ekrmv}
&\omega=e^{\rm spin}_{{\bm k}_{\parallel}-{\bm k}_{\parallel{\rm F}}^-}+\mu_-,\\
\label{eq:ekadd}
&\omega=-e^{\rm spin}_{-{\bm k}_{\parallel}+{\bm k}_{\parallel{\rm F}}^+}+\mu_+ 
\end{align}
(for $k_{\perp}=\pi$ and 0, respectively, in the Hubbard model). 
Here, 
\begin{align}
\label{eq:mu-}
\mu_-=E_{\rm GS}^{2N_{\rm u}}-E_{\rm GS}^{2N_{\rm u}-1},\\
\label{eq:mu+}
\mu_+=E_{\rm GS}^{2N_{\rm u}+1}-E_{\rm GS}^{2N_{\rm u}},
\end{align} 
which correspond to the $\omega$ values at the top of the lower band and bottom of the upper band 
in the zero-temperature electronic spectrum, respectively \cite{KohnoMottT}. 
When the Fermi level is located at the top of the lower band and bottom of the upper band, $\mu_-=0$ and $\mu_+=0$, 
which correspond to infinitesimal hole-doping and electron-doping, respectively (Secs. \ref{sec:holeDIS} and \ref{sec:eleDIS}). 
\par
This characteristic [Eqs. (\ref{eq:ekrmv}) and (\ref{eq:ekadd})] can be explained as follows \cite{KohnoMottT}: 
By adding an electron with a momentum of ${\bm k}_{\parallel}$ 
to the one-hole-doped ground state $|{\rm GS}\rangle_{\sigma^{\prime}}^{2N_{\rm u}-1}$ having a momentum of $-{\bm k}_{\parallel{\rm F}}^-$ involved in the thermal state, 
the obtained state has $N_{\rm e}=2N_{\rm u}$, a momentum of ${\bm k}_{\parallel}-{\bm k}_{\parallel{\rm F}}^-$, and a spin of $S=0$ or $1$. 
The $S=1$ state can overlap with the spin excited state at half filling 
$|{\cal T}({\bm k}_{\parallel}-{\bm k}_{\parallel{\rm F}}^-)\rangle$, as described in Sec. \ref{sec:holeDIS} [the second term of Eq. (\ref{eq:add})]. 
In the electronic spectrum, the dispersion relation of the emergent mode is obtained as Eq. (\ref{eq:ekrmv}), 
by putting $E_m=e^{\rm spin}_{{\bm k}_{\parallel}-{\bm k}_{\parallel{\rm F}}^-}+E_{\rm GS}^{2N_{\rm u}}$ and $E_n=E_{\rm GS}^{2N_{\rm u}-1}$ in Eq. (\ref{eq:Akw2}) 
[Eq. (\ref{eq:mu-})]. 
\par
Similarly, by removing an electron with a momentum of ${\bm k}_{\parallel}$ from the one-electron-doped ground state 
$|{\rm GS}\rangle_{\sigma^{\prime}}^{2N_{\rm u}+1}$ having a momentum of ${\bm k}_{\parallel{\rm F}}^+$ involved in the thermal state, 
the obtained state has $N_{\rm e}=2N_{\rm u}$, a momentum of $-{\bm k}_{\parallel}+{\bm k}_{\parallel{\rm F}}^+$, and a spin of $S=0$ or $1$. 
The $S=1$ state can overlap with the spin excited state at half filling $|{\cal T}(-{\bm k}_{\parallel}+{\bm k}_{\parallel{\rm F}}^+)\rangle$, 
as described in Sec. \ref{sec:eleDIS} [the second term of Eq. (\ref{eq:rmv})]. 
In the electronic spectrum, the dispersion relation of the emergent mode is obtained as Eq. (\ref{eq:ekadd}), 
by putting $E_m=E_{\rm GS}^{2N_{\rm u}+1}$ and $E_n=e^{\rm spin}_{-{\bm k}_{\parallel}+{\bm k}_{\parallel{\rm F}}^+}+E_{\rm GS}^{2N_{\rm u}}$ in Eq. (\ref{eq:Akw2}) 
[Eq. (\ref{eq:mu+})]. 
\par
In the case of $\beta e^{\rm spin}_{{\bm k}_{\parallel}}\lesssim 1$, the spin excited states can significantly contribute to the thermal state \cite{KohnoMottT}. 
According to Eq. (\ref{eq:Akw2}), electronic excitations from the spin excited states to the doped ground states, 
which correspond to the inverse processes of the above-described excitations \cite{KohnoMottT}, form 
electronic modes exhibiting the same dispersion relations as Eqs. (\ref{eq:ekrmv}) and (\ref{eq:ekadd}), 
by putting $|m\rangle=|{\cal T}({\bm k}_{\parallel}-{\bm k}_{\parallel{\rm F}}^-)\rangle$ and $|n\rangle=|{\rm GS}\rangle_{\sigma^{\prime}}^{2N_{\rm u}-1}$; 
$|m\rangle=|{\rm GS}\rangle_{\sigma^{\prime}}^{2N_{\rm u}+1}$ and $|n\rangle=|{\cal T}(-{\bm k}_{\parallel}+{\bm k}_{\parallel{\rm F}}^+)\rangle$ \cite{KohnoMottT}. 
\par
Thus, the temperature-induced in-gap modes exhibit essentially the same dispersion relations from the band edges 
as the hole- and electron-doping-induced in-gap modes [Eqs. (\ref{eq:ekHoleDIS}), (\ref{eq:ekEleDIS}), (\ref{eq:ekrmv}), and (\ref{eq:ekadd})] \cite{KohnoMottT}, 
as indicated by the solid red and magenta curves in Figs. \ref{fig:HubLadder}(d-3)--\ref{fig:HubLadder}(f-3), \ref{fig:HubBilayer}(d-3)--\ref{fig:HubBilayer}(f-3), 
\ref{fig:KLM}(c-3), and \ref{fig:KLM}(d-3), 
which explain the numerical results shown in Figs. \ref{fig:HubLadder}(a-3)--\ref{fig:HubLadder}(c-3), \ref{fig:HubBilayer}(a-3)--\ref{fig:HubBilayer}(c-3), 
\ref{fig:KLM}(a-3), and \ref{fig:KLM}(b-3). 
\subsubsection{Remarks on the spectral weight of temperature-induced states} 
\label{sec:remarks}
One of the reasons why the spectral weights of the emergent modes increase with temperature is explained as follows \cite{KohnoMottT}: 
When the temperature is comparable to the excitation energies 
($\beta e^{\rm spin}_{{\bm k}_{\parallel}}$, $\beta\varepsilon^{{\cal R},{\cal A}}_{{\bm k}_{\parallel}}\lesssim 1$), 
the spin and electronic excited states can have considerable Boltzmann weights and be significantly involved in the thermal state as 
$e^{-\frac{\beta E_{\rm GS}^{2N_{\rm u}}}{2}}e^{-\frac{\beta e^{\rm spin}_{{\bm k}_{\parallel}}}{2}}|{\cal T}({\bm k}_{\parallel})\rangle$, 
$e^{-\frac{\beta E_{\rm GS}^{2N_{\rm u}}}{2}}e^{-\frac{\beta\varepsilon^{{\cal R}}_{{\bm k}_{\parallel}}}{2}}|{\cal R}({\bm k}_{\parallel})\rangle$, and 
$e^{-\frac{\beta E_{\rm GS}^{2N_{\rm u}}}{2}}e^{-\frac{\beta\varepsilon^{{\cal A}}_{{\bm k}_{\parallel}}}{2}}|{\cal A}({\bm k}_{\parallel})\rangle$ [Eq. (\ref{eq:thermalState2})]. 
If the doped-ground-state components or spin-excited-state components are nonzero, electronic modes should emerge, exhibiting the dispersion relations 
shown in Eqs. (\ref{eq:ekrmv}) and (\ref{eq:ekadd}) (Sec. \ref{sec:TindModes}). 
The spectral weights of the emergent modes become larger as the components of the doped ground states or spin-excited states increase in the thermal state. 
\par
However, the spectral weight of the temperature-induced electronic state owing to the spin excited state $|{\cal T}({\bm k}_{\parallel}-{\bm k}_{\parallel{\rm F}}^-)\rangle$ or 
$|{\cal T}({-\bm k}_{\parallel}+{\bm k}_{\parallel{\rm F}}^+)\rangle$ is $\mathcal{O}(\frac{1}{N_{\rm u}})$, 
as in the case of the one-hole-doping- or one-electron-doping-induced state (Sec. \ref{sec:DIS}). 
According to the estimation of the spectral weight described in Sec. \ref{sec:holeDIS}, a macroscopic number [$\mathcal{O}(N_{\rm u})$] of holes, electrons, or spins 
should be excited to make the spectral weight comparable to those of zero-temperature bands; 
energies of such excited states are macroscopically higher than the ground-state energy $E_{\rm GS}^{2N_{\rm u}}$ 
[excitation energies from $|{\rm GS}\rangle^{2N_{\rm u}}$ are $\mathcal{O}(N_{\rm u}\varepsilon_{{\bm k}_{\parallel}}^{{\cal R},{\cal A}})$ 
or $\mathcal{O}(N_{\rm u}e^{\rm spin}_{{\bm k}_{\parallel}})$]. 
One might consider that the temperature comparable to the macroscopic excitation energies 
[$T\gtrsim\mathcal{O}(N_{\rm u}\varepsilon_{{\bm k}_{\parallel}}^{{\cal R},{\cal A}})$ or $\mathcal{O}(N_{\rm u}e^{\rm spin}_{{\bm k}_{\parallel}})$] would be required 
rather than $T\gtrsim\varepsilon_{{\bm k}_{\parallel}}^{{\cal R},{\cal A}}$ or $e^{\rm spin}_{{\bm k}_{\parallel}}$. 
\par
Nevertheless, the spectral weights of the emergent modes can become comparable to those of zero-temperature bands 
with temperature $T\gtrsim\varepsilon_{{\bm k}_{\parallel}}^{{\cal R},{\cal A}}$ or $e^{\rm spin}_{{\bm k}_{\parallel}}$, as shown in the numerical calculations \cite{KohnoMottT}. 
In the following sections, the reason is explained from the viewpoint of decoupled unit cells ($t=0$). 
\subsection{Temperature effects on decoupled unit cells} 
\label{sec:ucT}
\subsubsection{Selection rules and energies of electronic excitations} 
\label{sec:selectionrules}
To estimate the temperature scale for the robust emergent electronic modes, 
we first consider the $t=0$ case, where the system is decoupled into unit cells. 
Because the unit cells are independent of each other, the spectral function can be calculated only by considering a single unit cell. 
The partition function of $N_{\rm u}$ unit cells is simply $\Xi_1^{N_{\rm u}}$, where $\Xi_1$ denotes the partition function of a unit cell. 
\par
The spectral function [Eq. (\ref{eq:Akw2})] can be expressed as 
\begin{align}
A({\bm k},\omega)=\sum_{n,m}W_{mn}\delta(\omega-E_m+E_n) 
\label{eq:Akwuc}
\end{align}
using the spectral weight between states $|m\rangle$ and $|n\rangle$ in a unit cell defined as 
\begin{equation}
W_{mn}=\frac{1}{2\Xi_1}\sum_{\sigma}(e^{-\beta E_m}+e^{-\beta E_n})|\langle m|a_{\sigma}^{\dagger}|n\rangle|^2, 
\label{eq:WXY}
\end{equation}
where $E_n$ denotes the energy of $|n\rangle$ in a unit cell (Tables \ref{tbl:Hub} and \ref{tbl:KLM}), and 
$a_{\sigma}^{\dagger}$ denotes the creation operator of an electron with spin $\sigma$ in a unit cell [Eq. (\ref{eq:aidagger})]. 
Noted that $W_{mn}$ can be nonzero only when $N_{\rm e}$ and $S^z$ of $|m\rangle$ are larger than those of $|n\rangle$ by 1 and $s^z$, respectively. 
This is the selection rule of electronic excitations in a unit cell. 
Hereafter, the following shorthand notations are used: 
\begin{align}
&W_{XY}=\sum_{\sigma}W_{X_{\sigma}Y},&W_{YX}&=\sum_{\sigma}W_{YX_{\sigma}},\nonumber\\
&W_{{\rm T}X}=\sum_{\sigma,\gamma}W_{{\rm T}^{\gamma}X_{\sigma}},&W_{X{\rm T}}&=\sum_{\sigma,\gamma}W_{{X_{\sigma}\rm T}^{\gamma}},
\label{eq:WTY}
\end{align}
where $\gamma=1,0$, and $-1$; $X={\rm A}, {\rm B}, {\rm F}, {\rm G}, {\rm P}$, and ${\rm R}$; $Y=\psi^{+}, \psi^{-},{\rm D}^-, {\rm Z}, {\rm V}$, and ${\rm S}$. 
\par
\begin{figure*}
\includegraphics[width=\linewidth]{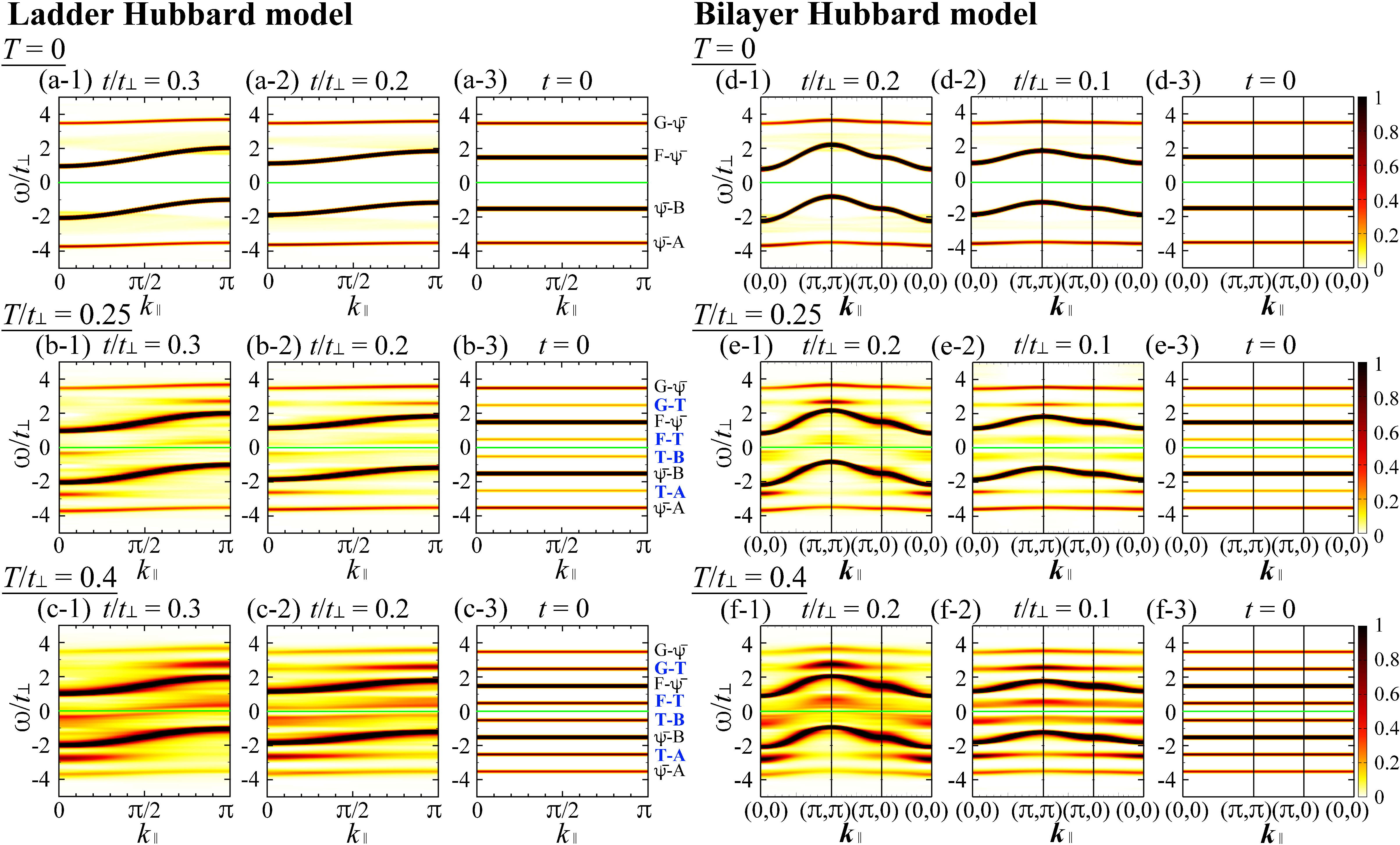}
\caption{Deformation of the band structure with inter-unit-cell hopping $t$ in the ladder and bilayer Hubbard models for $U/t_{\perp}=3$. 
(a-1)--(a-3), (b-1)--(b-3), (c-1)--(c-3) $A(k_{\parallel},0,\omega)t_{\perp}+A(k_{\parallel},\pi,\omega)t_{\perp}$ of the ladder Hubbard model; 
(d-1)--(d-3), (e-1)--(e-3), (f-1)--(f-3) ${\bar A}({\bm k}_{\parallel},0,\omega)t_{\perp}+{\bar A}({\bm k}_{\parallel},\pi,\omega)t_{\perp}$ 
of the bilayer Hubbard model at $T/t_{\perp}=0$ [(a-1)--(a-3), (d-1)--(d-3)], 0.25 [(b-1)--(b-3), (e-1)--(e-3)], and 0.4 [(c-1)--(c-3), (f-1)--(f-3)] 
for $t/t_{\perp}=0.3$ [(a-1)--(c-1)], 0.2 [(a-2)--(c-2), (d-1)--(f-1)], 0.1 [(d-2)--(f-2)], and 0 [(a-3)--(f-3)] 
obtained using the non-Abelian DDMRG method [(a-1), (a-2)] and CPT [(b-1)--(f-1), (b-2)--(f-2)]. 
Gaussian broadening with a standard deviation of $0.05t_{\perp}$ is used. The solid green lines indicate $\omega=0$. 
On the right side of (a-3)--(c-3), $X$-$Y$ indicates the mode with $\omega=E_X-E_Y$ for $X,Y=\psi^-, {\rm A}, {\rm B}, {\rm F}, {\rm G}$, and ${\rm T}$.}
\label{fig:tTdepHub}
\end{figure*}
\begin{figure*}
\includegraphics[width=\linewidth]{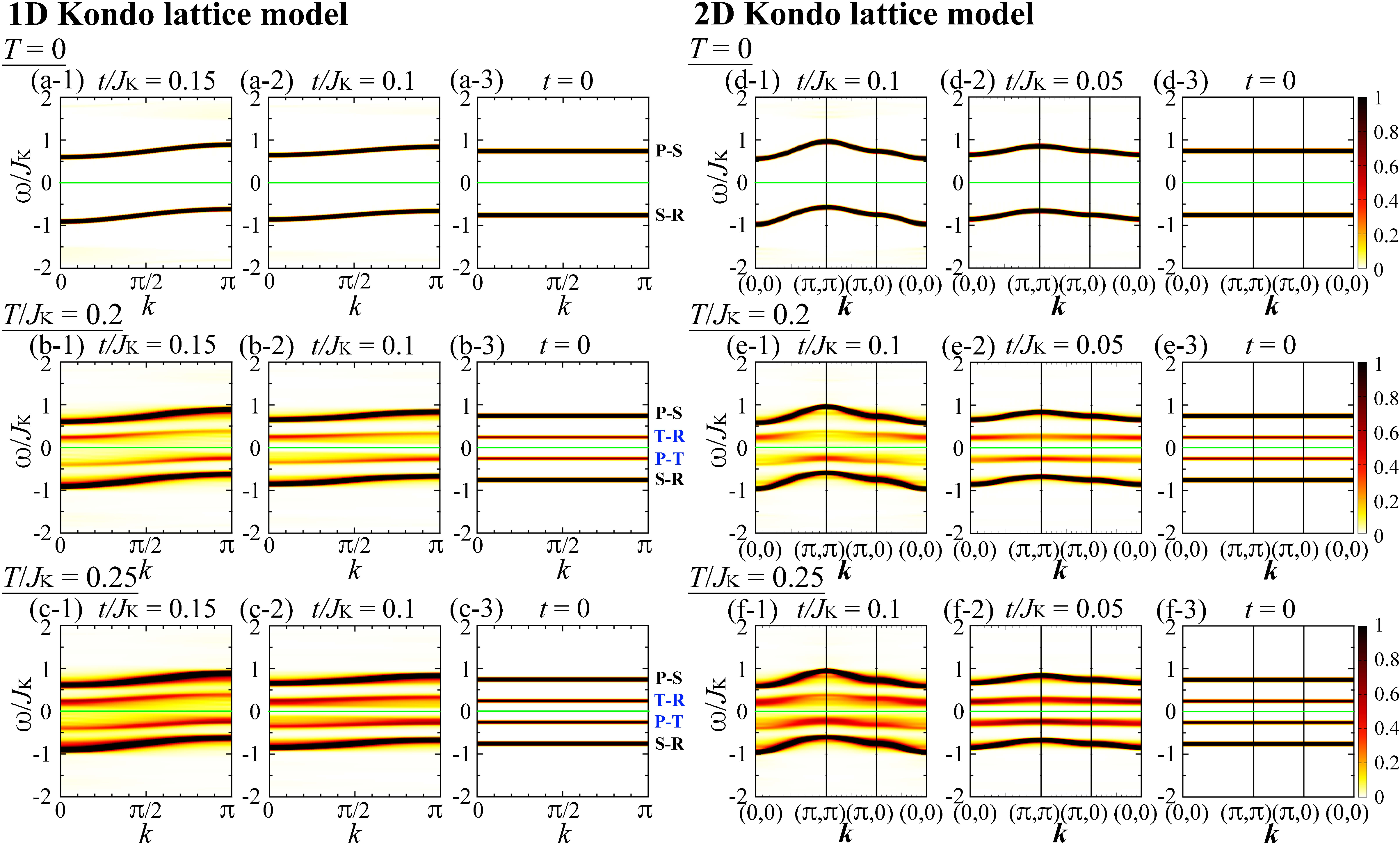}
\caption{Deformation of the band structure with inter-unit-cell hopping $t$ in the 1D and 2D KLMs. 
(a-1)--(a-3), (b-1)--(b-3), (c-1)--(c-3) $A(k,\omega)J_{\rm K}$ of the 1D KLM; (d-1)--(d-3), (e-1)--(e-3), (f-1)--(f-3) ${\bar A}({\bm k},\omega)J_{\rm K}$ of the 2D KLM 
at $T/J_{\rm K}=0$ [(a-1)--(a-3), (d-1)--(d-3)], 0.2 [(b-1)--(b-3), (e-1)--(e-3)], and 0.25 [(c-1)--(c-3), (f-1)--(f-3)] 
for $t/J_{\rm K}=0.15$ [(a-1)--(c-1)], 0.1 [(a-2)--(c-2), (d-1)--(f-1)], 0.05 [(d-2)--(f-2)], and 0 [(a-3)--(f-3)] 
obtained using the non-Abelian DDMRG method [(a-1), (a-2)] and CPT [(b-1)--(f-1), (b-2)--(f-2)]. 
Gaussian broadening with a standard deviation of $0.02J_{\rm K}$ is used. The solid green lines indicate $\omega=0$. 
On the right side of (a-3)--(c-3), $X$-$Y$ indicates the mode with $\omega=E_X-E_Y$ for $X,Y={\rm S}, {\rm P}, {\rm R}$, and ${\rm T}$.}
\label{fig:tTdepKLM}
\end{figure*}

Of note, the spectral function $A({\bm k},\omega)$ does not depend on momentum ${\bm k}_{\parallel}$ at $t=0$: 
the dispersion relations are flat as a function of momentum ${\bm k}_{\parallel}$ 
[Figs. \ref{fig:tTdepHub}(a-3)--\ref{fig:tTdepHub}(f-3) and \ref{fig:tTdepKLM}(a-3)--\ref{fig:tTdepKLM}(f-3)]. 
\par
At zero temperature, $A({\bm k},\omega)$ is nonzero at $\omega=-E_{\tilde {\cal R}}+E_{\cal G}$ and $\omega=E_{\tilde {\cal A}}-E_{\cal G}$ 
[$W_{\psi^-{\rm A}}$, $W_{\psi^-{\rm B}}$, $W_{{\rm F}\psi^-}$, $W_{{\rm G}\psi^-}\ne 0$ in the Hubbard model; 
$W_{\rm SR}$, $W_{\rm PS}\ne 0$ in the KLM (Tables \ref{tbl:Hub}, \ref{tbl:KLM}, and \ref{tbl:symbols})]. 
In the Hubbard model, the excited states $|{\rm B}_{\sigma}\rangle$ and $|{\rm F}_{\sigma}\rangle$ have significant spectral weights for $\omega<0$ at $k_{\perp}=0$ 
and for $\omega>0$ at $k_{\perp}=\pi$, respectively [Figs. \ref{fig:tTdepHub}(a-3) and \ref{fig:tTdepHub}(d-3)]. 
In the KLM, $|{\rm R}_{\sigma}\rangle$ and $|{\rm P}_{\sigma}\rangle$ have spectral weights for $\omega<0$ and $\omega>0$, respectively 
[Figs. \ref{fig:tTdepKLM}(a-3) and \ref{fig:tTdepKLM}(d-3)]. 
\par
\begin{figure*}
\includegraphics[width=\linewidth]{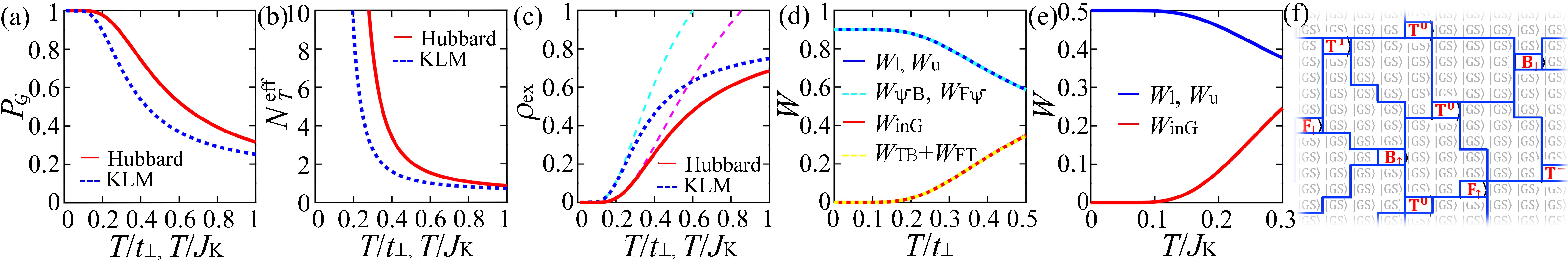}
\caption{Temperature effects on unit cells. 
(a) Probability of the ground state $P_{\cal G}$($=$ the density of ground-state unit cells in a system of decoupled unit cells), 
(b) typical size of an effective cluster undisturbed by temperature $N^{\rm eff}_T$, 
(c) density of excited unit cells $\rho_{\rm ex}$ for the $U/t_{\perp}=3$ Hubbard model as a function of $T/t_{\perp}$ (solid red curve) and 
the KLM as a function of $T/J_{\rm K}$ (dotted blue curve). 
In (c), $1/N^{\rm eff}_T$ is shown for the $U/t_{\perp}=3$ Hubbard model (dashed magenta curve) and the KLM (dashed cyan curve). 
(d) Spectral weight of the dimer Hubbard model for $U/t_{\perp}=3$ in the lower dominant band $W_{\rm l}(=W_{\psi^-{\rm B}}+W_{{\rm G}\psi^+})$ and 
the upper dominant band $W_{\rm u}(=W_{{\rm F}\psi^-}+W_{\psi^+{\rm A}})$ (solid blue curve) and 
that in the in-gap bands $W_{\rm inG}(=W_{\rm TB}+W_{\rm FT}+W_{{\rm G}{\rm D}^-}+W_{{\rm D}^-{\rm A}}+W_{\rm AV}+W_{\rm ZG})$ (solid red curve). 
The dominant parts of the lower and upper dominant bands are $W_{\psi^-{\rm B}}$ and $W_{{\rm F}\psi^-}$, respectively (dotted cyan curve), 
and those of the in-gap bands are $W_{\rm TB}$ and $W_{\rm FT}$ (dotted yellow curve). 
(e) Spectral weight of the single-site KLM in the lower band $W_{\rm l}(=W_{\rm SR})$ and the upper band $W_{\rm u}(=W_{\rm PS})$ (solid blue curve) 
and that in the in-gap bands $W_{\rm inG}(=W_{\rm TR}+W_{\rm PT})$ (solid red curve). 
(f) Image of a typical configuration in a system of decoupled unit cells; 
the system consists of effective clusters undisturbed by temperature (regions enclosed by blue lines) and excited unit cells between them (red states).}
\label{fig:dimer}
\end{figure*}

At nonzero temperature, spectral weights emerge at 
\begin{equation}
\begin{array}{l}
\omega=E_{\rm T}-E_{\cal R}=e^{\rm spin}_{{\bm k}_{\parallel}-{\bm k}_{\parallel{\rm F}}^-}+\mu_-,\\
\omega=-E_{\rm T}+E_{\cal A}=-e^{\rm spin}_{-{\bm k}_{\parallel}+{\bm k}_{\parallel{\rm F}}^+}+\mu_+
\end{array}
\end{equation}
(for $k_{\perp}=\pi$ and 0, respectively, in the Hubbard model), 
where $\mu_-=E_{\cal G}-E_{\cal R}$, $\mu_+=E_{\cal A}-E_{\cal G}$, and $e^{\rm spin}_{{\bm k}_{\parallel}}=E_{\rm T}-E_{\cal G}$ 
[Eqs. (\ref{eq:spinExE}), (\ref{eq:mu-}), and (\ref{eq:mu+}) for $t=0$]; 
$e^{\rm spin}_{{\bm k}_{\parallel}}$ does not depend on momentum ${\bm k}_{\parallel}$ at $t=0$. 
These are the temperature-induced modes within the band gap 
[Figs. \ref{fig:tTdepHub}(b-3), \ref{fig:tTdepHub}(e-3), \ref{fig:tTdepKLM}(b-3), and \ref{fig:tTdepKLM}(e-3)]. 
As the temperature increases, the spectral weights of the emergent modes increase 
[Figs. \ref{fig:tTdepHub}(c-3), \ref{fig:tTdepHub}(f-3), \ref{fig:tTdepKLM}(c-3), and \ref{fig:tTdepKLM}(f-3); Figs. \ref{fig:dimer}(d) and \ref{fig:dimer}(e)] 
and become comparable to those of zero-temperature bands 
with temperature $T\gtrsim e^{\rm spin}_{{\bm k}_{\parallel}}$ or $\varepsilon^{{\cal R},{\cal A}}_{{\bm k}_{\parallel}}$. 
\par
In the Hubbard model, the high-$|\omega|$ modes are also induced and grow with temperature at 
\begin{equation}
\begin{array}{lllll}
\omega=E_{\rm G}-E_{\rm T}&{\rm for}&\omega>0&{\rm at}&k_{\perp}=\pi,\\
\omega=E_{\rm T}-E_{\rm A}&{\rm for}&\omega<0&{\rm at}&k_{\perp}=0
\end{array}
\end{equation}
[Figs. \ref{fig:tTdepHub}(b-3), \ref{fig:tTdepHub}(c-3), \ref{fig:tTdepHub}(e-3), and \ref{fig:tTdepHub}(f-3)]. 
\par
The spectral weights of the emergent modes at $t=0$ are determined by the Boltzmann weights and matrix elements of the states in a unit cell 
[Eq. (\ref{eq:WXY})], where the norms of the matrix elements are $\mathcal{O}(1)$ independent of temperature. 
The spectral weight of the electronic excitation between $|{\cal T}\rangle$ and $|{\cal R}\rangle$ or $|{\cal A}\rangle$ can be substantial in each unit cell 
if the Boltzmann weights are comparable to that of the ground state $|{\cal G}\rangle$ in a unit cell 
($\beta e^{\rm spin}_{{\bm k}_{\parallel}}, \beta\varepsilon^{{\cal R},{\cal A}}_{{\bm k}_{\parallel}}\lesssim 1$). 
Because the spectral weights of the flat modes at each momentum ${\bm k}_{\parallel}$ are the same as those in a unit cell for $t=0$ [Eqs. (\ref{eq:Akwuc}) and (\ref{eq:WXY})], 
the spectral weights of the emergent modes can be comparable to those of zero-temperature bands 
at temperature $T\gtrsim e^{\rm spin}_{{\bm k}_{\parallel}}$ or $\varepsilon^{{\cal R},{\cal A}}_{{\bm k}_{\parallel}}$. 
\par
Notably, these results are exact even in the bulk limit ($N_{\rm u}\rightarrow\infty$); 
the temperature required to make the spectral weight of the temperature-induced states considerable is 
not the excitation energy from the ground state $|{\rm GS}\rangle^{2N_{\rm u}}$ but the spin or electronic excitation energies in a unit cell. 
At the temperature where the temperature-induced states have considerable spectral weights, the thermal state involves a macroscopic number of excited unit cells 
(Sec. \ref{sec:holeDIS}), and the excitation energies in a unit cell ($e^{\rm spin}_{{\bm k}_{\parallel}}$ and $\varepsilon^{{\cal R},{\cal A}}_{{\bm k}_{\parallel}}$) are nonzero. 
Hence, the excitation energy of the thermal state from the ground state $|{\rm GS}\rangle^{2N_{\rm u}}$ 
is macroscopic: $\mathcal{O}(N_{\rm u}e^{\rm spin}_{{\bm k}_{\parallel}})$ or $\mathcal{O}(N_{\rm u}\varepsilon^{{\cal R},{\cal A}}_{{\bm k}_{\parallel}})$ (Sec. \ref{sec:remarks}). 
Nevertheless, the temperature required to make the spectral weight of temperature-induced states considerable is 
$\mathcal{O}(e^{\rm spin}_{{\bm k}_{\parallel}})$ or $\mathcal{O}(\varepsilon^{{\cal R},{\cal A}}_{{\bm k}_{\parallel}})$. 
This is because temperature is an intensive thermodynamic quantity that acts on all the unit cells simultaneously. 
\subsubsection{Effective clusters undisturbed by temperature} 
\label{sec:thermalCluster}
Because the spectral weights of emergent modes are basically proportional to the density of excited unit cells (Sec. \ref{sec:holeDIS}), 
we consider the temperature dependence of the density of excited unit cells at $t=0$. 
The probability of state $|n\rangle$ with energy $E_n$ in a unit cell is obtained as 
\begin{equation}
P_{n}=\frac{e^{-\beta E_n}}{\Xi_1}, 
\end{equation}
where $\Xi_1$ denotes the partition function in a unit cell: $\Xi_1=\sum_m e^{-\beta E_m}$. 
At zero temperature, all the unit cells are in the ground state $|{\cal G}\rangle$ ($P_{\cal G}=1$, $P_{n\ne{\cal G}}=0$); 
the ground state of $N_{\rm u}$ unit cells is $|{\rm GS}\rangle^{2N_{\rm u}}$ [Eq. (\ref{eq:GS})] at half filling. 
The probability that $m$ unit cells around a unit cell remain in the ground state is expressed as 
\begin{equation}
\label{eq:PG}
\begin{array}{lll}
({P_{\cal G}})^m=e^{-\frac{m}{N^{\rm eff}_T}},&{\rm where}&N^{\rm eff}_T=-\frac{1}{\ln{P_{\cal G}}}. 
\end{array}
\end{equation}
This implies that the unit cells within an effective cluster of size $N^{\rm eff}_T$ basically remain in the ground state. 
Thus, the system can be regarded as an ensemble of effective clusters undisturbed by temperature, whose typical size is $N^{\rm eff}_T$, 
and excited unit cells with density $\rho_{\rm ex}=1-P_{\cal G}$ [Fig. \ref{fig:dimer}(f)]. 
Because excited unit cells typically appear in every $N^{\rm eff}_T$ unit cells, 
the number of excited unit cells can be estimated as $\frac{N_{\rm u}}{N^{\rm eff}_T}$ in the $N_{\rm u}$-unit-cell system in the low-temperature regime. 
In fact, 
\begin{equation}
\frac{1}{N^{\rm eff}_T}\overset{{\rm Eq.} (\ref{eq:PG})}{=}-\ln{P_{\cal G}}=-\ln({1-\rho_{\rm ex})}\overset{\rho_{\rm ex}\ll1}{\approx}\rho_{\rm ex}.
\label{eq:rhoex}
\end{equation}
\par
In Fig. \ref{fig:dimer}, the probability of the ground state in a unit cell $P_{\cal G}$ [Fig. \ref{fig:dimer}(a)], 
the typical size of an undisturbed cluster $N^{\rm eff}_T$ [Fig. \ref{fig:dimer}(b)], and the density of excited unit cells 
$\rho_{\rm ex}(\approx\frac{1}{N^{\rm eff}_T}$ for $\rho_{\rm ex}\ll1$) [Fig. \ref{fig:dimer}(c)] 
as a function of temperature are shown for the $U/t_{\perp}=3$ Hubbard model and the KLM. 
\par
In the low-temperature regime of $T\ll\Delta$, where $\Delta$ denotes the lowest excitation energy in a unit cell 
($\Delta=t_{\perp}$ for the $U/t_{\perp}=3$ Hubbard model; $\Delta=\frac{3}{4}J_{\rm K}$ for the KLM at $\mu=0$), 
almost all the unit cells are in the ground state: $P_{\cal G}\approx 1$ [Fig. \ref{fig:dimer}(a)]. 
The typical size of an undisturbed cluster $N^{\rm eff}_T$ is very large ($N^{\rm eff}_T\gg 1$) [Fig. \ref{fig:dimer}(b)], and 
the density of excited unit cells $\rho_{\rm ex}$ is very small ($\rho_{\rm ex}\approx\frac{1}{N^{\rm eff}_T}\approx 0$) [Fig. \ref{fig:dimer}(c)]. 
This implies that stable states at temperatures $T\ll\Delta$ are almost the ground state; there are almost no excited unit cells. 
\par
As the temperature increases above $T\approx 0.2\Delta$, 
the probability of the ground-state unit cell $P_{\cal G}$ decreases substantially [Fig. \ref{fig:dimer}(a)]. 
The typical size of an undisturbed cluster $N^{\rm eff}_T$ quickly decreases and becomes as large as several unit cells [Fig. \ref{fig:dimer}(b)], 
which partly justifies numerical calculations in small-size clusters to investigate spectral features at temperatures comparable to typical excitation energies, 
and the density of excited unit cells $\rho_{\rm ex}$ becomes considerable [Fig. \ref{fig:dimer}(c)]. 
The spectral weights of the emergent states also start to increase substantially at $T\approx 0.2\Delta$ [Figs. \ref{fig:dimer}(d) and \ref{fig:dimer}(e)]. 
\par
\subsection{Interpretation of temperature-induced states based on undisturbed clusters} 
When the spectral weights of temperature-induced modes become comparable to those of zero-temperature bands, 
the number of excited unit cells are macroscopic: $\mathcal{O}(N_{\rm u})$ (Sec. \ref{sec:holeDIS}). 
One might consider that the states relevant to temperature-induced electronic excitations would not be related to spin excitation at zero temperature 
because their energies differ macroscopically [$\mathcal{O}(N_{\rm u}\rho_{\rm ex}\Delta)$ at $t=0$ (Sec. \ref{sec:thermalCluster})]. 
\par
Nevertheless, the essence of spin excitation at zero temperature is relevant to the temperature-induced electronic states. 
At $t=0$, the spin-excitation energy in a unit cell is the same as that of the bulk at zero temperature, 
$e^{\rm spin}_{{\bm k}_{\parallel}}=E_{\rm T}-E_{\cal G}$ [Eq. (\ref{eq:spinExE})]. 
Even with a small $|t|$, the spin-excitation energy does not change much. 
In the low-temperature regime, the spin-excitation energy in an undisturbed cluster is expected to be essentially the same as that of the bulk at zero temperature 
because the undisturbed cluster is defined within which all the unit cells remain in the ground state; 
the spin-excitation dispersion relation in an undisturbed cluster remains essentially the same as that of the bulk, $\omega\approx e^{\rm spin}_{{\bm k}_{\parallel}}$. 
As discussed in Sec. \ref{sec:TindStates}, temperature-induced states emerge when the thermal state involves an excited particle (hole, electron, or spin). 
Hence, in an undisturbed cluster with a thermally excited particle, by adding and removing an electron, temperature-induced states emerge with dispersion relations 
essentially the same as those of the bulk [Sec. \ref{sec:TindModes}; Eqs. (\ref{eq:ekrmv}) and (\ref{eq:ekadd})]. 
\par
As for the spectral weight, the spectral weight of the temperature-induced states in each undisturbed cluster with a thermally excited particle is 
$\mathcal{O}(\frac{1}{N_{\rm u}})$ for $a_{{\bm k}_{\parallel},\sigma}^{(\dagger)}$ (Sec. \ref{sec:remarks}). 
The number of such clusters can be estimated as $\frac{N_{\rm u}}{(N^{\rm eff}_T+1)}\approx\frac{N_{\rm u}}{N^{\rm eff}_T}\approx N_{\rm u}\rho_{\rm ex}$ 
in the low-temperature regime [Eq. (\ref{eq:rhoex})]. 
Hence, the total spectral weight of the temperature-induced states at momentum ${\bm k}_{\parallel}$ is $\mathcal{O}(\rho_{\rm ex})$. 
\par
Thus, by regarding the spin excitation as that in an undisturbed cluster, the explanations for the temperature-induced modes 
[Sec. \ref{sec:TindStates}; Eqs. (\ref{eq:ekrmv}) and (\ref{eq:ekadd})] \cite{KohnoMottT} are basically valid, 
and the physical picture should persist up to temperatures comparable to the spin-excitation energies, where the temperature-induced modes carry considerable spectral weights. 
Hence, the temperature-induced modes originating from spin excitation [Sec. \ref{sec:TindStates}; Eqs. (\ref{eq:ekrmv}) and (\ref{eq:ekadd})] \cite{KohnoMottT} 
can gain spectral weights comparable to those of zero-temperature bands 
when the temperature increases up to about the spin-excitation energies ($T\gtrsim e^{\rm spin}_{{\bm k}_{\parallel}}$) 
[Figs. \ref{fig:wt}(a-2)--\ref{fig:wt}(d-2), \ref{fig:tTdepHub}(c-1)--\ref{fig:tTdepHub}(c-3), \ref{fig:tTdepHub}(f-1)--\ref{fig:tTdepHub}(f-3), 
\ref{fig:tTdepKLM}(c-1)--\ref{fig:tTdepKLM}(c-3), \ref{fig:tTdepKLM}(f-1)--\ref{fig:tTdepKLM}(f-3), \ref{fig:dimer}(d), and \ref{fig:dimer}(e)]. 
The above argument is generally applicable to spin-gapped strongly correlated insulators, regardless of the lattice structure. 
\subsection{Effects of inter-unit-cell hopping on the band structure} 
At $t=0$, the spectral weights of temperature-induced modes can become comparable to those of zero-temperature bands 
for $T\gtrsim e^{\rm spin}_{{\bm k}_{\parallel}}$ or $\varepsilon^{{\cal R},{\cal A}}_{{\bm k}_{\parallel}}$ 
[Sec. \ref{sec:ucT}; Figs. \ref{fig:tTdepHub}(c-3), \ref{fig:tTdepHub}(f-3), \ref{fig:tTdepKLM}(c-3), \ref{fig:tTdepKLM}(f-3), \ref{fig:dimer}(d), and \ref{fig:dimer}(e)]. 
As $|t|$ increases, the spectral-weight distribution continuously changes from that of $t=0$; 
the temperature-induced modes gradually become dispersing (Figs. \ref{fig:tTdepHub} and \ref{fig:tTdepKLM}). 
Their spectral weights do not decrease but rather increase with $|t|$ [Figs. \ref{fig:wt}(a-3)--\ref{fig:wt}(d-3) and \ref{fig:wt}(a-4)--\ref{fig:wt}(d-4)]; 
the spectral weights of emergent modes are expected to increase with $|t|$ because $|t|$ lowers the band bottom of the gapped excitation, 
which increases the Boltzmann weight of the band bottom, usually increasing the spectral weight of the spin mode around the bottom. 
\par
Thus, the spectral weights of temperature-induced modes generally remain comparable to those of zero-temperature bands 
even with considerable inter-unit-cell hopping $|t|$ for $T\gtrsim e^{\rm spin}_{{\bm k}_{\parallel}}$ or $\varepsilon^{{\cal R},{\cal A}}_{{\bm k}_{\parallel}}$. 
This implies that temperature-induced modes can generally be regarded as bands with considerable spectral weights and 
change the band structure from that of zero temperature, 
provided that the temperature increases up to about the spin-excitation energies ($T\gtrsim e^{\rm spin}_{{\bm k}_{\parallel}}$).
\par
In spin-gapped Mott and Kondo insulators, the spectral weights of temperature-induced states increase exponentially with temperature 
in the low-temperature regime [Figs. \ref{fig:wt}(a-2)--\ref{fig:wt}(d-2), \ref{fig:dimer}(d), and \ref{fig:dimer}(e)] \cite{KohnoKLM}, reflecting the gapped excitation. 
This is in contrast to the doping evolution of the spectral weight of the doping-induced states, which is proportional to $|\delta|$ in the small-doping regime 
[Figs. \ref{fig:wt}(a-1)--\ref{fig:wt}(d-1); Sec. \ref{sec:DIS}]. 
\par
Based on the continuous change in the spectral-weight distribution with $t$ (Figs. \ref{fig:tTdepHub} and \ref{fig:tTdepKLM}), 
the in-gap modes of the ladder and bilayer Hubbard models can be identified as originating primarily 
from the excitations of $|{\rm F}_{\sigma}\rangle\leftrightarrow|{\cal T}\rangle$ and $|{\cal T}\rangle\leftrightarrow|{\rm B}_{\sigma}\rangle$ [Fig. \ref{fig:dimer}(d)], 
and the high-$|\omega|$ emergent modes as $|{\rm G}_{\sigma}\rangle\leftrightarrow|{\cal T}\rangle$ and $|{\cal T}\rangle\leftrightarrow|{\rm A}_{\sigma}\rangle$. 
The in-gap modes of the 1D and 2D KLMs can be identified as originating from the excitations of $|{\cal T}\rangle\leftrightarrow|{\rm R}_{\sigma}\rangle$ and 
$|{\rm P}_{\sigma}\rangle\leftrightarrow|{\cal T}\rangle$. 
Hence, the temperature-induced electronic modes can be basically understood as excitations related to the spin excited state $|{\cal T}\rangle$. 
\subsection{Effective theory at nonzero temperature} 
\label{sec:AkwT}
\subsubsection{Noninteracting excited-particle approximation} 
To discuss the dispersion relations of temperature-induced electronic modes, 
we make an approximation where the excited particles in the effective theory for weak inter-unit-cell hopping are assumed to be noninteracting. 
This approximation is justified in the dilute limit of excited particles (low-temperature limit) where excited particles can essentially move freely. 
In this approximation, the spectral weight is overestimated and the sum rule is not satisfied, but the essential spectral features can be captured as shown below. 
\par
In the noninteracting excited-particle approximation, the temperature-induced particles $X$ are assumed to move freely without recognizing the existence of other particles, 
where $X={\rm T}^{\gamma}, {\rm D}^-, {\rm V}, {\rm Z}, {\rm A}_{\sigma}, {\rm B}_{\sigma}, {\rm F}_{\sigma}$, and ${\rm G}_{\sigma}$ in the Hubbard model (Table \ref{tbl:Hub}); 
$X={\rm T}^{\gamma}, {\rm P}_{\sigma}$, and ${\rm R}_{\sigma}$ in the KLM (Table \ref{tbl:KLM}) for $\gamma=1,0,-1$ and $\sigma=\uparrow, \downarrow$. 
The single-particle state of $X$ with momentum ${\bm k}_{\parallel}$ is expressed as Eq. (\ref{eq:Xk}). 
The partition function can be factorized with respect to excited particles as $\Xi=\prod_X\Xi_X$, and 
the partition function of each particle can further be factorized with respect to momentum ${\bm k}_{\parallel}$ as $\Xi_X=\prod_{{\bm k}_{\parallel}}\Xi_{X_{{\bm k}_{\parallel}}}$. 
The numbers of bosons (${\rm T}^{\gamma}$, ${\rm D}^-$, ${\rm V}$, and ${\rm Z}$ in the Hubbard model; ${\rm T}^{\gamma}$ in the KLM) 
and fermions (${\rm A}_{\sigma}, {\rm B}_{\sigma}, {\rm F}_{\sigma}$, and ${\rm G}_{\sigma}$ in the Hubbard model; 
${\rm P}_{\sigma}$ and ${\rm R}_{\sigma}$ in the KLM) with momentum ${\bm k}_{\parallel}$ can be expressed as Bose and Fermi distribution functions, respectively. 
The distribution functions of bosons ($-$) and fermions ($+$) with momentum ${\bm k}_{\parallel}$ are defined as 
\begin{equation}
\label{eq:distfun}
f^{\pm}(\epsilon^X_{{\bm k}_{\parallel}})=\frac{1}{e^{\beta\epsilon^X_{{\bm k}_{\parallel}}}\pm1}, 
\end{equation}
where $\epsilon^X_{{\bm k}_{\parallel}}$ denotes the excitation energy from the ground state; 
$\epsilon^{X}_{{\bm k}_{\parallel}}=\varepsilon^{X}_{{\bm k}_{\parallel}}$ for fermion $X_{\sigma}$ [Eq. (\ref{eq:eleExE})], 
$e^{\rm spin}_{{\bm k}_{\parallel}}$ for $X={\rm T}^{\gamma}$ [Eq. (\ref{eq:spinExE})], 
and $e^{X}_{{\bm k}_{\parallel}}$ for $X={\rm D}^-$, ${\rm V}$, and ${\rm Z}$ [Eq. (\ref{eq:chargeExE})]. 
\par
Thus, the spectral functions [Eq. (\ref{eq:Akw2})] for the emergent electronic states are approximated as 
\begin{align}
\label{eq:AkwTX}
A^{X-{\rm T}}({\bm k},\omega)=&\frac{1}{2}
\sum_{\sigma,{\bm p}_{\parallel}}[f^-(e^{\rm spin}_{-{\bm k}_{\parallel}+{\bm p}_{\parallel}})+f^+(\varepsilon^X_{{\bm p}_{\parallel}})]\nonumber\\
&\times[|\langle {\rm T}^{-2s^z}(-{\bm k}_{\parallel}+{\bm p}_{\parallel})|a_{{\bm k}_{\parallel},\sigma}|X_{\bar \sigma}({\bm p}_{\parallel})\rangle|^2\nonumber\\
&+|\langle {\rm T}^0(-{\bm k}_{\parallel}+{\bm p}_{\parallel})|a_{{\bm k}_{\parallel},\sigma}|X_{\sigma}({\bm p}_{\parallel})\rangle|^2]\nonumber\\
&\times\delta(\omega+e^{\rm spin}_{-{\bm k}_{\parallel}+{\bm p}_{\parallel}}-\varepsilon^X_{{\bm p}_{\parallel}}),
\end{align}
\begin{align}
\label{eq:AkwTY}
A^{{\rm T}-Y}({\bm k},\omega)=&\frac{1}{2}
\sum_{\sigma,{\bm p}_{\parallel}}[f^-(e^{\rm spin}_{{\bm k}_{\parallel}+{\bm p}_{\parallel}})+f^+(\varepsilon^Y_{{\bm p}_{\parallel}})]\nonumber\\
&\times[|\langle {\rm T}^{2s^z}({\bm k}_{\parallel}+{\bm p}_{\parallel})|a_{{\bm k}_{\parallel},\sigma}^{\dagger}|Y_{\sigma}({\bm p}_{\parallel})\rangle|^2\nonumber\\
&+|\langle {\rm T}^0({\bm k}_{\parallel}+{\bm p}_{\parallel})|a_{{\bm k}_{\parallel},\sigma}^{\dagger}|Y_{\bar \sigma}({\bm p}_{\parallel})\rangle|^2]\nonumber\\
&\times\delta(\omega-e^{\rm spin}_{{\bm k}_{\parallel}+{\bm p}_{\parallel}}+\varepsilon^Y_{{\bm p}_{\parallel}}),
\end{align}
where $k_{\perp}=0$ for $X={\rm F}$ and $Y={\rm A}$; $k_{\perp}=\pi$ for $X={\rm G}$ and $Y={\rm B}$ in the Hubbard model, 
and ${\bm k}={\bm k}_{\parallel}$ for $X={\rm P}$ and $Y={\rm R}$ in the KLM. 
\par
By considering the first-order correction to the single-particle state [state without correction is the right-hand side of Eq. (\ref{eq:Xk})], 
the squared norms of the matrix elements in Eqs. (\ref{eq:AkwTX}) and (\ref{eq:AkwTY}) are obtained up to $\mathcal{O}(t)$ as 
\begin{align}
\label{eq:elmTX}
&|\langle {\rm T}^{-2s^z}(-{\bm k}_{\parallel}+{\bm p}_{\parallel})|a_{{\bm k}_{\parallel},\sigma}|X_{\bar \sigma}({\bm p}_{\parallel})\rangle|^2\nonumber\\
&=2|\langle {\rm T}^0(-{\bm k}_{\parallel}+{\bm p}_{\parallel})|a_{{\bm k}_{\parallel},\sigma}|X_{\sigma}({\bm p}_{\parallel})\rangle|^2\nonumber\\
&=\frac{1}{N_{\rm u}}\left[1-\frac{2tdv_{-\eta}^2\gamma_{{\bm k}_{\parallel}}}{E^-+\frac{U}{2}-2\eta t_{\perp}}
-\frac{2tdv_{\eta}^2(\gamma_{{\bm p}_{\parallel}}-\gamma_{{\bm k}_{\parallel}})}{E^--\frac{U}{2}+2\eta t_{\perp}}\right],\\
\label{eq:elmTY}
&|\langle {\rm T}^{2s^z}({\bm k}_{\parallel}+{\bm p}_{\parallel})|a_{{\bm k}_{\parallel},\sigma}^{\dagger}|Y_{\sigma}({\bm p}_{\parallel})\rangle|^2\nonumber\\
&=2|\langle {\rm T}^0({\bm k}_{\parallel}+{\bm p}_{\parallel})|a_{{\bm k}_{\parallel},\sigma}^{\dagger}|Y_{\bar \sigma}({\bm p}_{\parallel})\rangle|^2\nonumber\\
&=\frac{1}{N_{\rm u}}\left[1+\frac{2tdv_{-\eta}^2\gamma_{{\bm k}_{\parallel}}}{E^-+\frac{U}{2}-2\eta t_{\perp}}
+\frac{2tdv_{\eta}^2(\gamma_{{\bm p}_{\parallel}}-\gamma_{{\bm k}_{\parallel}})}{E^--\frac{U}{2}+2\eta t_{\perp}}\right]
\end{align}
in the Hubbard model, where $v_{\pm 1}^2=1\pm\frac{4t_{\perp}}{\sqrt{U^2+16t^2}}$; 
$\eta=1$ for $X={\rm F}$ and $Y={\rm B}$; $\eta=-1$ for $X={\rm G}$ and $Y={\rm A}$. 
In the KLM, 
\begin{align}
\label{eq:elmTA}
&|\langle {\rm T}^{-2s^z}(-{\bm k}_{\parallel}+{\bm p}_{\parallel})|a_{{\bm k}_{\parallel},\sigma}|{\rm P}_{\bar \sigma}({\bm p}_{\parallel})\rangle|^2\nonumber\\
&=2|\langle {\rm T}^0(-{\bm k}_{\parallel}+{\bm p}_{\parallel})|a_{{\bm k}_{\parallel},\sigma}|{\rm P}_{\sigma}({\bm p}_{\parallel})\rangle|^2\nonumber\\
&=\frac{1}{N_{\rm u}}\left[1+\frac{2td(2\gamma_{{\bm p}_{\parallel}}-\gamma_{{\bm k}_{\parallel}})}{J_{\rm K}}\right],\\
\label{eq:elmTR}
&|\langle {\rm T}^{2s^z}({\bm k}_{\parallel}+{\bm p}_{\parallel})|a^{\dagger}_{{\bm k}_{\parallel},\sigma}|{\rm R}_{\sigma}({\bm p}_{\parallel})\rangle|^2\nonumber\\
&=2|\langle {\rm T}^0({\bm k}_{\parallel}+{\bm p}_{\parallel})|a^{\dagger}_{{\bm k}_{\parallel},\sigma}|{\rm R}_{\bar \sigma}({\bm p}_{\parallel})\rangle|^2\nonumber\\
&=\frac{1}{N_{\rm u}}\left[1-\frac{2td(2\gamma_{{\bm p}_{\parallel}}-\gamma_{{\bm k}_{\parallel}})}{J_{\rm K}}\right].
\end{align}
\par
\begin{figure*}
\includegraphics[width=\linewidth]{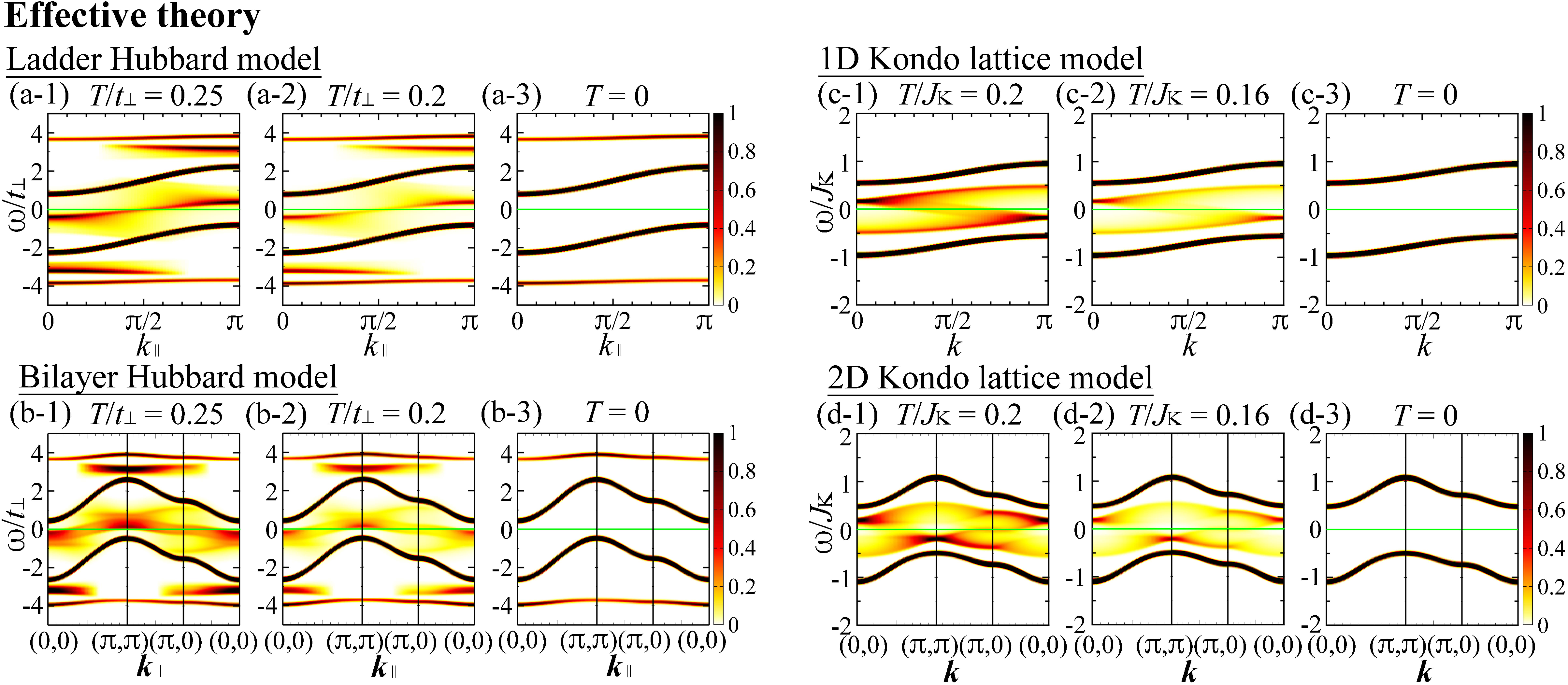}
\caption{Evolution of temperature-induced electronic bands obtained using the effective theory for weak inter-unit-cell hopping 
with noninteracting excited-particle approximation. 
(a-1)--(a-3) $A(k_{\parallel},0,\omega)t_{\perp}+A(k_{\parallel},\pi,\omega)t_{\perp}$ of the ladder Hubbard model for $t/t_{\perp}=0.4$ and $U/t_{\perp}=3$ 
at $T/t_{\perp}=0.25$ [(a-1)], 0.2 [(a-2)], and 0 [(a-3)]. 
(b-1)--(b-3) $A({\bm k}_{\parallel},0,\omega)t_{\perp}+A({\bm k}_{\parallel},\pi,\omega)t_{\perp}$ of the bilayer Hubbard model for $t/t_{\perp}=0.3$ and $U/t_{\perp}=3$ 
at $T/t_{\perp}=0.25$ [(b-1)], 0.2 [(b-2)], and 0 [(b-3)]. 
(c-1)--(c-3) $A(k,\omega)J_{\rm K}$ of the 1D KLM for $t/J_{\rm K}=0.2$ at $T/J_{\rm K}=0.2$ [(c-1)], 0.16 [(c-2)], and 0 [(c-3)]. 
(d-1)--(d-3) $A({\bm k},\omega)J_{\rm K}$ of the 2D KLM for $t/J_{\rm K}=0.15$ at $T/J_{\rm K}=0.2$ [(d-1)], 0.16 [(d-2)], and 0 [(d-3)]. 
Gaussian broadening is used with a standard deviation of $0.05t_{\perp}$ [(a-1)--(a-3), (b-1)--(b-3)] and $0.02J_{\rm K}$ [(c-1)--(c-3), (d-1)--(d-3)]. 
The solid green lines indicate $\omega=0$.}
\label{fig:effAkwT}
\end{figure*}

Figure \ref{fig:effAkwT} shows the spectral functions obtained in this approximation for the emergent states [Eqs. (\ref{eq:AkwTX}) and (\ref{eq:AkwTY})], 
the ${\rm A}$, ${\rm B}$, ${\rm F}$, and ${\rm G}$ modes in the ladder and bilayer Hubbard models, and the ${\rm P}$ and ${\rm R}$ modes in the 1D and 2D KLMs 
with corrections up to $\mathcal{O}(t)$ in the squared norms of the matrix elements. 
The spectral features at nonzero temperatures [Figs. \ref{fig:HubLadder}(a-3), \ref{fig:HubBilayer}(a-3), \ref{fig:KLM}(a-3), and \ref{fig:KLM}(b-3)] are captured reasonably well, 
although the spectral weight is overestimated in this approximation. 
\subsubsection{Extraction of temperature-induced modes} 
Notably, the squared norm of the matrix element between $|{\cal T}(-{\bm k}_{\parallel}+{\bm p}_{\parallel})\rangle$ and 
the electron-added state [$|{\rm F}_{\sigma}({\bm p}_{\parallel})\rangle$ and $|{\rm G}_{\sigma}({\bm p}_{\parallel})\rangle$ in the Hubbard model; 
$|{\rm P}_{\sigma}({\bm p}_{\parallel})\rangle$ in the KLM] 
becomes the largest at ${\bm p}_{\parallel}={\bm 0}(={\bm k}_{\parallel {\rm F}}^+)$ for $t>0$ at each ${\bm k}_{\parallel}$ 
[Eq. (\ref{eq:elmTX}), where $E^--\frac{U}{2}\pm 2t_{\perp}<0$ and $v_{\pm 1}^2>0$, and Eq. (\ref{eq:elmTA}); Eq. (\ref{eq:gammak})], 
and that between $|{\cal T}({\bm k}_{\parallel}+{\bm p}_{\parallel})\rangle$ and the electron-removed state 
[$|{\rm B}_{\sigma}({\bm p}_{\parallel})\rangle$ and $|{\rm A}_{\sigma}({\bm p}_{\parallel})\rangle$ in the Hubbard model; 
$|{\rm R}_{\sigma}({\bm p}_{\parallel})\rangle$ in the KLM] becomes the largest 
at ${\bm p}_{\parallel}=-{\bm \pi}(=-{\bm k}_{\parallel {\rm F}}^-)$ for $t>0$ at each ${\bm k}_{\parallel}$ 
[Eqs. (\ref{eq:elmTY}) and (\ref{eq:elmTR})]. 
\par
This means that the dominant modes in the emergent electronic spectra can be identified as those excited from the band edges: 
the bottom of the upper band [$|{\rm F}_{\sigma}({\bm k}_{\parallel {\rm F}}^+)\rangle$ and $|{\rm G}_{\sigma}({\bm k}_{\parallel {\rm F}}^+)\rangle$ in the Hubbard model; 
$|{\rm P}_{\sigma}({\bm k}_{\parallel {\rm F}}^+)\rangle$ in the KLM] 
and the top of the lower band [$|{\rm B}_{\sigma}(-{\bm k}_{\parallel {\rm F}}^-)\rangle$ and $|{\rm A}_{\sigma}(-{\bm k}_{\parallel {\rm F}}^-)\rangle$ in the Hubbard model; 
$|{\rm R}_{\sigma}(-{\bm k}_{\parallel {\rm F}}^-)\rangle$ in the KLM]. 
\par
In fact, by approximating the dominant parts of Eqs. (\ref{eq:elmTX})--(\ref{eq:elmTR}) as 
\begin{align}
\label{eq:elmTXeff}
&|\langle {\rm T}^{-2s^z}(-{\bm k}_{\parallel}+{\bm p}_{\parallel})|a_{{\bm k}_{\parallel},\sigma}|X_{\bar \sigma}({\bm p}_{\parallel})\rangle|^2\nonumber\\
&=2|\langle {\rm T}^0(-{\bm k}_{\parallel}+{\bm p}_{\parallel})|a_{{\bm k}_{\parallel},\sigma}|X_{\sigma}({\bm p}_{\parallel})\rangle|^2\nonumber\\
&\approx C_{{\bm k}_{\parallel}}^+\delta_{{\bm p}_{\parallel},{\bm k}_{{\parallel}{\rm F}}^+},\\
\label{eq:elmTYeff}
&|\langle {\rm T}^{2s^z}({\bm k}_{\parallel}+{\bm p}_{\parallel})|a_{{\bm k}_{\parallel},\sigma}^{\dagger}|Y_{\sigma}({\bm p}_{\parallel})\rangle|^2\nonumber\\
&=2|\langle {\rm T}^0({\bm k}_{\parallel}+{\bm p}_{\parallel})|a_{{\bm k}_{\parallel},\sigma}^{\dagger}|Y_{\bar \sigma}({\bm p}_{\parallel})\rangle|^2\nonumber\\
&\approx C_{{\bm k}_{\parallel}}^-\delta_{{\bm p}_{\parallel},-{\bm k}_{{\parallel}{\rm F}}^-}
\end{align}
for $X={\rm F}$ at $k_{\perp}=0$ and $Y={\rm B}$ at $k_{\perp}=\pi$ in the Hubbard model; $X={\rm P}$ and $Y={\rm R}$ in the KLM 
and by plugging them into Eqs. (\ref{eq:AkwTX}) and (\ref{eq:AkwTY}), the dispersion relations of the dominant modes are extracted as 
\begin{align}
\label{eq:ekaddeff}
\omega=&-e^{\rm spin}_{-{\bm k}_{\parallel}+{\bm k}_{{\parallel}{\rm F}}^+}+\varepsilon^X_{{\bm k}_{{\parallel}{\rm F}}^+},\\
\label{eq:ekrmveff}
\omega=&e^{\rm spin}_{{\bm k}_{\parallel}-{\bm k}_{{\parallel}{\rm F}}^-}-\varepsilon^Y_{-{\bm k}_{{\parallel}{\rm F}}^-}
\end{align}
(for $k_{\perp}=0$ and $\pi$, respectively, in the Hubbard model). 
Equations (\ref{eq:ekaddeff}) and (\ref{eq:ekrmveff}) reduce to Eqs. (\ref{eq:ekadd}) and (\ref{eq:ekrmv}), respectively, because 
\begin{equation}
\begin{array}{ll}
\varepsilon^X_{{\bm k}_{{\parallel}{\rm F}}^+}&
\overset{{\rm Eq.} (\ref{eq:1eleEGS})}{=}E_{\rm GS}^{2N_{\rm u}+1}-E_{\rm GS}^{2N_{\rm u}}\overset{{\rm Eq.} (\ref{eq:mu+})}{=}\mu_+,\\
\varepsilon^Y_{-{\bm k}_{{\parallel}{\rm F}}^-}&
\overset{{\rm Eq.} (\ref{eq:1holeEGS})}{=}E_{\rm GS}^{2N_{\rm u}-1}-E_{\rm GS}^{2N_{\rm u}}\overset{{\rm Eq.} (\ref{eq:mu-})}{=}-\mu_-.
\end{array}
\end{equation}
\par
Thus, the dominant electronic modes induced by temperature can be identified as originating from the spin-excitation mode, 
which exhibit the spin-mode dispersion relations shifted by the Fermi momenta (${\bm k}_{\parallel{\rm F}}^{\pm}$) from the band edges ($\mu_{\pm}$) 
[Eqs. (\ref{eq:ekrmv}) and (\ref{eq:ekadd})] \cite{KohnoMottT}. 
This feature is generally expected in Mott and Kondo insulators \cite{KohnoMottT}. 
\par
In the ladder and bilayer Hubbard models, electronic modes are induced by temperature not only within the band gap but also in the high-$|\omega|$ regimes. 
The dispersion relations are given by Eq. (\ref{eq:ekaddeff}) with $X={\rm G}$ at $k_{\perp}=\pi$ for $\omega>0$ 
[dashed red curves in Figs. \ref{fig:HubLadder}(d-3), \ref{fig:HubLadder}(f-3), \ref{fig:HubBilayer}(d-3), and \ref{fig:HubBilayer}(f-3); 
Figs. \ref{fig:HubLadder}(a-3), \ref{fig:HubLadder}(c-3), \ref{fig:HubBilayer}(a-3), and \ref{fig:HubBilayer}(c-3)] 
and by Eq. (\ref{eq:ekrmveff}) with $Y={\rm A}$ at $k_{\perp}=0$ for $\omega<0$ 
[dashed magenta curves in Figs. \ref{fig:HubLadder}(d-3), \ref{fig:HubLadder}(e-3), \ref{fig:HubBilayer}(d-3), and \ref{fig:HubBilayer}(e-3); 
Figs. \ref{fig:HubLadder}(a-3), \ref{fig:HubLadder}(b-3), \ref{fig:HubBilayer}(a-3), and \ref{fig:HubBilayer}(b-3)], 
\begin{equation}
\begin{array}{lll}
\omega=-e^{\rm spin}_{-{\bm k}_{\parallel}+{\bm k}_{{\parallel}{\rm F}}^+}+\varepsilon^{\rm G}_{{\bm k}_{{\parallel}{\rm F}}^+}&{\rm for}&k_{\perp}=\pi,\\
\omega=e^{\rm spin}_{{\bm k}_{\parallel}-{\bm k}_{{\parallel}{\rm F}}^-}-\varepsilon^{\rm A}_{-{\bm k}_{{\parallel}{\rm F}}^-}&{\rm for}&k_{\perp}=0.
\end{array}
\label{eq:ekhighwT}
\end{equation}
Note that the dispersion relations of the temperature-induced high-$|\omega|$ modes differ from those of the doping-induced high-$|\omega|$ modes 
[Eqs. (\ref{eq:ekD-}), (\ref{eq:ekV}), and (\ref{eq:ekZ})]. 
\par
The spectral weights of temperature-induced states [Eqs. (\ref{eq:AkwTX}) and (\ref{eq:AkwTY})] increase with temperature 
because the values of distribution functions of the excited particles [Eq. (\ref{eq:distfun})] increase with temperature 
in the noninteracting excited-particle approximation (Fig. \ref{fig:effAkwT}). 
The temperature required to make the spectral weights of temperature-induced states considerable is 
of the order of the excitation energies $\mathcal{O}(\epsilon^X_{{\bm k}_{\parallel}})$ 
rather than $\mathcal{O}(N_{\rm u}\epsilon^X_{{\bm k}_{\parallel}})$. 
This is because temperature is an intensive thermodynamic quantity that acts on all the excited particles for all the momenta simultaneously. 
\subsection{Spin-charge separation of strongly correlated insulators} 
\label{sec:spinchargeSep}
The spin and charge dynamical structure factors generally differ from each other in interacting-electron systems in any spatial dimension, and 
the spin and charge degrees of freedom are generally coupled with each other in the energy regime away from the low excitation-energy limit 
in interacting metals in any spatial dimension. 
\par
Spin-charge separation means that the lowest spin-excitation energy differs from the lowest charge-excitation energy. 
In an interacting metal on a chain, it is known that the spin and charge velocities, 
i.e., the lowest spin- and charge-excitation energies of $\mathcal{O}(\frac{1}{L})$, are generally different, 
where $L$ denotes the number of sites on a chain \cite{Essler,HaldaneTLL,TomonagaTLL,LuttingerTLL,MattisLiebTLL,Giamarchi}. 
This is the spin-charge separation usually considered as a characteristic of 1D interacting metals, 
which is contrasted with a Fermi liquid where spin and charge excitations are obtained as particle-hole excitations of electronic quasiparticles \cite{LandauFL}; 
hence, the lowest spin- and charge-excitation energies are the same. 
\par
\begin{figure*}
\includegraphics[width=\linewidth]{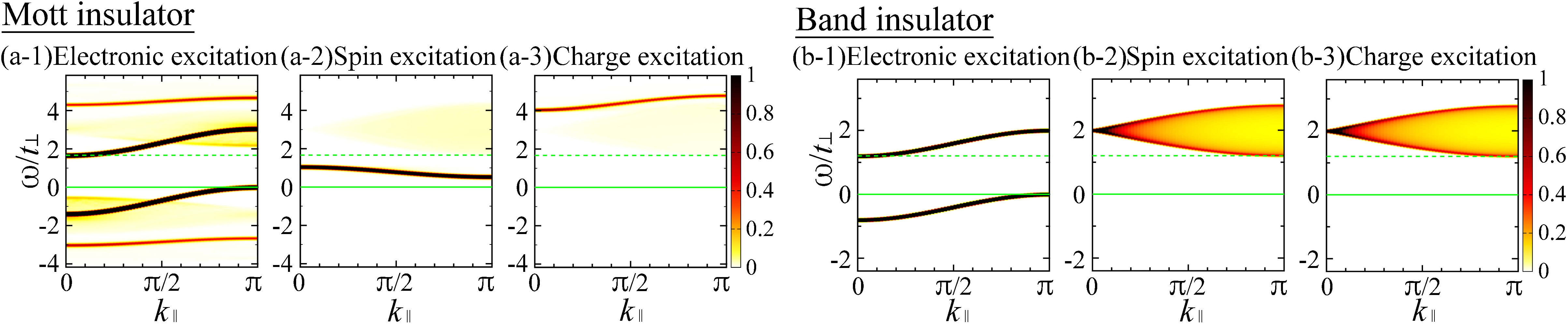}
\caption{Spin-charge separation of a Mott insulator [(a-1)--(a-3)] and non-spin-charge separation of a band insulator [(b-1)--(b-3)]. 
(a-1) $A(k_{\parallel},0,\omega)t_{\perp}+A(k_{\parallel},\pi,\omega)t_{\perp}$, (a-2) $S_{\rm Hub}(k_{\parallel},\omega)t_{\perp}/3$, 
(a-3) $N_{\rm Hub}(k_{\parallel},\omega)t_{\perp}/4$ of the ladder Hubbard model at half filling for $U/t_{\perp}=3$ and $t/t_{\perp}=0.4$ 
obtained using the non-Abelian DDMRG method. 
(b-1) $A(k_{\parallel},0,\omega)t_{\perp}+A(k_{\parallel},\pi,\omega)t_{\perp}$, (b-2) $S_{\rm Hub}(k_{\parallel},\omega)t_{\perp}/3$, 
(b-3) $N_{\rm Hub}(k_{\parallel},\omega)t_{\perp}/4$ of the ladder Hubbard model at half filling for $U=0$ and $t/t_{\perp}=0.2$. 
The chemical potential $\mu$ is adjusted such that the Fermi level is located at the top of the lower band. 
Gaussian broadening is used with a standard deviation of $0.05t_{\perp}$ [(a-1)--(a-3)] and $0.02t_{\perp}$ [(b-1)--(b-3)]. 
The solid green lines indicate $\omega=0$. The dotted green lines indicate the bottom of the upper band.}
\label{fig:U0ladder}
\end{figure*}

In strongly correlated insulators such as Mott and Kondo insulators, spin-charge separation generally occurs in any spatial dimension. 
For example, in the large-$U/t$ Hubbard model, the energy of spin excitation described by the Heisenberg model is $\mathcal{O}(J=\frac{4t^2}{U})$, 
which is much smaller than the lowest charge-excitation energy of $\mathcal{O}(U)$ \cite{Essler,Anderson,TakahashiBook,LiebWu}; 
the lowest spin- and charge-excitation energies are different. 
Specifically, as shown in Figs. \ref{fig:U0ladder}(a-1)--\ref{fig:U0ladder}(a-3), the spin excitation exists in the energy regime within the band gap, 
\begin{equation}
0<e^{\rm spin}_{{\bm k}_{\parallel}}<\mu_+-\mu_-.
\end{equation}
The charge-excitation mode appears above the continuum [Fig. \ref{fig:U0ladder}(a-3)], which corresponds to the ${\rm D}^-$ mode 
exhibiting $\omega=e^{{\rm D}^-}_{{\bm k}_{\parallel}}$ [Eq. (\ref{eq:chargeExE})]. 
\par
By contrast, in a band insulator described as noninteracting electrons on a lattice, spin-charge separation does not occur in any spatial dimension; 
both the lowest spin- and charge-excitation energies are equal to the band gap [Figs. \ref{fig:U0ladder}(b-1)--\ref{fig:U0ladder}(b-3)]. 
In fact, the lowest-energy spin excited state where the $z$ component of spin is raised by one is obtained 
by removing a down-spin electron from the top of the filled lower band ($c_{\downarrow,{\bm k}_{\rm F}^-}$) and 
adding an up-spin electron to the bottom of the empty upper band ($c^{\dagger}_{\uparrow,{\bm k}_{\rm F}^+}$). 
This excitation energy is equal to the band gap. 
Similarly, the lowest charge-excitation energy is also equal to the band gap. 
Hence, spin-charge separation does not occur in a band insulator. 
This is because the lowest-energy spin and charge excitations are constructed as particle-hole excitations of electrons in a band insulator, as in a Fermi liquid. 
Similarly, in the mean-field theory and density-functional theory where the Hamiltonian is expressed as noninteracting electronic quasiparticles in effective potential, 
spin-charge separation does not occur. 
To explain the emergent electronic modes reflecting the spin excitation beyond the mean-field theory, spin fluctuation corresponding to the spin excitation 
should be considered in the electronic self-energy, as shown in Ref. \cite{KohnoAF}. 
\par
The emergence of electronic modes exhibiting momentum-shifted magnetic dispersion relations from the band edges 
within the band gap by temperature and doping discussed in this paper as well as 
in Refs. [\onlinecite{KohnoMottT,KohnoRPP,Kohno1DHub,Kohno2DHub,Kohno1DtJ,Kohno2DtJ,KohnoDIS,KohnoHubLadder,KohnoSpin,KohnoAF,KohnoGW,KohnoKLM,
KohnottpHub,KohnottpJ}] 
reflects spin-charge separation of Mott and Kondo insulators, 
i.e., the existence of spin excited states whose excitation energies are lower than the charge gap (band gap), regardless of the spatial dimension. 
\subsection{Implications} 
\label{sec:implications}
In this paper, the spectral features of spin-gapped Mott and Kondo insulators were explained from the viewpoint of weak inter-unit-cell hopping. 
As $|t|$ increases, details of the dispersion relations and spectral-weight distributions might be substantially affected by higher-order corrections of the perturbation theory. 
Nevertheless, the essential features presented in this paper should be valid as long as the spin gap opens. 
Even if spin excitation is gapless, electronic modes exhibiting momentum-shifted magnetic dispersion relations should emerge from the band edges and evolve 
with temperature and doping, according to the quantum-number analyses and numerical calculations 
\cite{KohnoMottT,KohnoRPP,Kohno1DHub,Kohno2DHub,Kohno1DtJ,Kohno2DtJ,KohnoSpin,KohnoAF,KohnoGW}; 
this reflects spin-charge separation of strongly correlated insulators (Sec. \ref{sec:spinchargeSep}; Fig. \ref{fig:U0ladder}). 
Furthermore, for gapless spin excitation, the spectral weights of temperature-induced electronic modes should increase more rapidly with temperature than the spin-gapped case. 
\par
Thus, the temperature-induced electronic modes essentially exhibiting momentum-shifted magnetic dispersion relations from the band edges 
should be observed in various models and materials of strongly correlated insulators, regardless of the spatial dimension, 
quasiparticle picture (with or without quasiparticle weight at the Fermi momentum in a doped system), or the presence or absence of a spin gap or antiferromagnetic order 
\cite{KohnoMottT,KohnoRPP,Kohno1DHub,Kohno2DHub,Kohno1DtJ,Kohno2DtJ,KohnoDIS,KohnoHubLadder,KohnoSpin,KohnoAF,KohnoGW,KohnoKLM}, 
provided that the temperature increases up to about spin-excitation energies. 
\par
Although various interpretations on electronic states in strongly correlated systems have been considered 
\cite{KohnoMottT,KohnoRPP,Kohno1DHub,Kohno2DHub,Kohno1DtJ,Kohno2DtJ,KohnoDIS,KohnoHubLadder,KohnoSpin,KohnoAF,KohnoGW,KohnoKLM,KohnottpHub,KohnottpJ,Eskes,DagottoDOS,NoceraDMRT_T,Matsueda_QMC,PreussQP,GroberQMC,VCALanczos,TPQSLanczos,KuzminCPT,PAMMPS,HubbardI,TsunetsuguRMP,OneSiteKondo,Essler,HaldaneTLL,TomonagaTLL,LuttingerTLL,MattisLiebTLL,Giamarchi,LandauFL,BrinkmanRice,ImadaRMP,SakaiImadaPRL,SakaiImadaPRB,ImadaCofermionPRL,ImadaCofermionPRB,PhillipsRMP,PhillipsRPP,EderOhta2DHub,EderOhtaIPES}, 
the view presented in this paper and 
Refs. [\onlinecite{KohnoMottT,KohnoRPP,Kohno1DHub,Kohno2DHub,Kohno1DtJ,Kohno2DtJ,KohnoDIS,KohnoHubLadder,KohnoSpin,KohnoAF,KohnoGW,KohnoKLM,
KohnottpHub,KohnottpJ}] can consistently explain the characteristic temperature- and doping-induced electronic states around Mott and Kondo insulators quite generally 
and clarify the distinction from the temperature- and doping-independent band structure of band insulators. 
According to this view, the number of electronic bands increases by temperature and doping if spin excitation has an energy gap in Mott and Kondo insulators 
\cite{KohnoMottT,KohnoDIS,KohnoHubLadder,KohnoGW,KohnoKLM}; 
if spin excitation is gapless, gapless temperature- and doping-induced electronic modes emerge from the band edges 
\cite{KohnoMottT,KohnoRPP,Kohno1DHub,Kohno2DHub,Kohno1DtJ,Kohno2DtJ,KohnoSpin,KohnoAF,KohnoGW,KohnottpHub,KohnottpJ}. 
If the temperature-induced modes cross the Fermi level, the band structure can be regarded as metallic \cite{KohnoMottT}. 
\section{Summary} 
The origin of temperature-induced modes, their relation to doping-induced modes, and how their spectral weights increase with temperature 
were clarified from the viewpoint of weakly coupled unit cells for spin-gapped Mott and Kondo insulators. 
The temperature evolution of emergent electronic states was demonstrated in the case of decoupled unit cells, 
where the spectral function can be calculated exactly at any temperature. 
By introducing effective clusters undisturbed by temperature, within which all the unit cells remain in the ground state, 
this study explained that the temperature-induced modes originating from spin excitation can generally gain spectral weights 
comparable to those of zero-temperature bands, 
provided that the temperature increases up to about the spin-excitation energies. 
Furthermore, by applying the effective theory for weak inter-unit-cell hopping to the nonzero-temperature case with noninteracting excited-particle approximation, 
the emergence and evolution of electronic states with temperature were demonstrated, 
and the dominant temperature-induced modes were shown to exhibit momentum-shifted magnetic dispersion relations from the band edges. 
Numerical results for the ladder and bilayer Hubbard models and 1D and 2D KLMs were explained in detail using the effective theory. 
\par
As shown in this paper, the temperature-induced electronic modes originating from spin excitation can generally grow into robust bands 
as the temperature increases up to about the spin-excitation energies, even without a phase transition or lattice distortion. 
In spin-gapped Mott and Kondo insulators, the number of bands increases by temperature as well as by doping, because the emergent in-gap modes can be regarded as bands. 
This feature reflects spin-charge separation, i.e., the existence of spin excitation in the energy regime lower than the charge gap, 
which is general and fundamental in strongly correlated insulators, regardless of the spatial dimension. 
\par
Although the band structure of spin-gapped Mott and Kondo insulators is sometimes difficult to distinguish from that of band insulators, 
the emergence and evolution of electronic modes exhibiting momentum-shifted magnetic dispersion relations from the band edges 
can be observed by increasing the temperature without the need for doping, 
which can be used as a principle of band-gap engineering in strong-correlation electronics. 
\begin{acknowledgments} 
This work was supported by JSPS KAKENHI Grant No. JP22K03477 and World Premier International Research Center Initiative (WPI), MEXT, Japan. 
Numerical calculations were partly performed on the Numerical Materials Simulator at NIMS. 
\end{acknowledgments}

\end{document}